# Spatial multiplexing of atom-photon entanglement sources using feed-forward control and switching networks

*Long Tian, Zhongxiao Xu, Lirong Chen, Wei Ge, Haoxiang Yuan, Yafei Wen, Shengzhi Wang,*

*Shujing Li and Hai Wang\**

*The State Key Laboratory of Quantum Optics and Quantum Optics Devices,*

*Collaborative Innovation Center of Extreme Optics,*

*Institute of Opto-Electronics, Shanxi University, Taiyuan, 030006,*

*People's Republic of China*

The light-matter quantum interface that can create quantum-correlations or entanglement between a photon and one atomic collective excitation is a fundamental building block for a quantum repeater. The intrinsic limit is that the probability of preparing such non-classical atom-photon correlations has to be kept low in order to suppress multi-excitation. To enhance this probability without introducing multi-excitation errors, a promising scheme is to apply multimode memories into the interface. Significant progress has been made in temporal, spectral, and spatial multiplexing memories, but the enhanced probability for generating the entangled atom-photon pair has not been experimentally realized. Here, by using six spin-wave-photon entanglement sources, a switching network and feed-forward control, we build a multiplexed light-matter interface and then demonstrate a ~six-fold (~four-fold) probability increase in generating entangled atom-photon (photon-photon) pairs. The measured compositive Bell parameter for the multiplexed interface is 2.49±0.03 combined with a memory lifetime of up to ~51μs.

\*corresponding author：wanghai@sxu.edu.cn

The distribution of entanglement over a long distance is a central task of quantum communication [1-5]. Due to transmission losses in optical fibers, directly distributing entanglement over long distance (>500km) is limited [6-8]. For overcoming this limitation, Briegel *et al.* proposed a quantum repeater (QR) protocol [9], in which the entanglement distance between two locations is divided into $N$ shorter elementary links. Entanglement is generated in each link and then successively extended via entanglement swapping between two adjacent links. To realize a practical QR, an attractive approach is the DLCZ protocol [3], which uses atomic ensembles as memory elements and single-photon detection for entanglement creation and swapping. Based on the DLCZ protocol, several improved QR schemes have been proposed [10-15], in which, a robust QR scheme [10-11] has been paid more attention. In this scheme, entanglement swapping and connection are achieved by using two-photon detection, thereby the long-distance phase stability, required in DLCZ protocol, is no longer necessary [1]. The fundamental building block for the DLCZ-type or robust QR protocols [1, 10-11] is a light-mater quantum interface (LMQI) that can generate entanglement between a photon and an atomic collective excitation in a probabilistic way. Such LMQIs have been experimentally demonstrated by using spontaneous Raman scattering in cold atomic ensembles [16-30], or storing one of correlated or entangled photons in a solid-state [31-34] or gas-state atomic ensemble [35]. Additionally, entanglement between a single photon and a single quantum-system such as a NV-center [36], or a trapped atom/ion [37-41], has been experimentally demonstrated and proposed for alternative QR approaches [42-45]. The atomic-ensemble based LMQIs [16-35] are attractive because they use relatively simple ingredients [1, 14]. However, the probability to prepare the atom-photon entanglement in such LMQIs has to be kept low for avoiding multi-excitation errors. So, the entanglement distribution over a long distance requires a long storage time, e.g., 1000km entanglement distribution requires a time of 10-second order, which exceeds the state-of-the-art results [21]. For greatly reducing the required time, the multiplexed QR protocols are proposed [46, 47], in which, the temporally or spatially multiplexed memories are introduced into the atomic-ensemble based LMQI and the generation probability of atom-photon correlation or entanglement pairs will be greatly enhanced [1, 46, 47]. In recent years, many multimode-memory experiments, including storage of 5 temporally-multiplexed polarization qubits [48] and 26 spectrally-multiplexed time-bin qubits [49] at the single-photon-level, as well as 50 temporally-multiplexed light pulses [50] have been demonstrated. It is worth noting that the authors in Ref. [49] utilized feed-forward-controlled operations on the retrieved qubits encoded into the multiplexed spectral modes. The LMQI with six spectral-modes [51], and with 12-independently-addressable spatial-mode memories [52] have been demonstrated and the atom-photon correlation and entanglement were observed in the two experiments, respectively. Recently, Tiranov *et al.* demonstrated and certified the simultaneous storage and retrieval of two entangled photons inside a solid-state quantum memory [53]. However, the enhanced preparation rate for the atom-photon correlated or entangled pair has not been observed in these experiments due to lack of feed-forward-controlled write (storing) [52] or read (retrieving) [51, 53] process.

Here, we demonstrate a multiplexed light-matter-entanglement interface (LMEI) formed by a spatial array consisting of six spin-wave-photon entanglement (SWPE) sources together with a feed-forward-controlled optical switching network (OSN) and then achieve a six-fold (four-fold) enhancement of the atom-photon (photon-photon) entanglement generation probability without introducing extra noise. In addition, also the compositive fidelity (Bell parameter) is used for characterizing the quality of the entanglement created in the multiplexed LMEI, whose measured maximal value is 87.9%±1% (2.49±0.03), overcoming the critical limit of 78% [54]. In contrast to the previous spatially-multiplexed schemes [47, 52, 55], where multiple independent sub-ensembles are used as memory elements, our presented scheme uses only multiple spatial modes in an ensemble as memory elements.

We now show that our multiplexed LMEI is available for the QR protocol shown in Fig.1, where, $m$ SWPE sources generated from an atomic ensemble are located at the end point of each elementary link (i.e., node). In an elementary link, e.g., A-B link in Fig.1(a), a single photon coming from the $i$th source located at the left (A) ensemble is sent to the $i$th central station (CS$_i$) of the link to meet another photon coming from the $i$th source from the right (B) ensemble. At each of the stations CS$_i$ ($i=1,...,m$), the two photons will undergo a Bell-state measurement (BSM) behind a polarization-beam-splitter (PBS). A successful BSM, e.g., at the $k$th station (CS$_k$), indicates that an entangled state between two spin-wave qubits, which are stored in $k$th spatial modes of the left and right ensembles, respectively, is created, i.e. entanglement between the two ensembles is established. For performing the entanglement swapping between the A-B and C-D links [Fig.1(b)], one converts the spin-wave excitations stored in the B and C ensembles into two anti-Stokes photons, respectively, and combines them on a PBS. Since the retrieved photon from the B (C) ensemble may come from any of the spatial modes, one has to route it into a common channel (see Sec. IV in Supplemental Material [56] for details) for achieving this swapping.

Aiming at the above multiplexed QR scheme, we experimentally demonstrate a multiplexed interface (MI) by using six SWPE sources in a single atomic ensemble utilizing feed-forward control. The experimental setup and relevant atomic levels are shown in Fig. 2. The atomic ensemble is a cloud of cold $^{87}$Rb atoms, whose two ground levels $|a\rangle$ and $|b\rangle$, together with the excited level $|e_1\rangle$ ($|e_2\rangle$), forms a $\Lambda$-type configuration. The atoms are prepared in the initial state $|a\rangle$ and then an off-resonant $\sigma^+$ – polarized writing pulse is applied onto them along z-axis. The writing pulses induce

Raman transition $|a\rangle \to |b\rangle$ via $|e_1\rangle$ [Fig. 2(b)], which emits Stokes photons and simultaneously creates spin-wave excitations. We collect the Stokes photons in different spatial modes $S_i$ ($i=1,...,6$) at small angles relative to the z-axis. If a single Stokes photon is emitted in the spatial mode $S_i$, one atomic spin-wave excitation will be created and stored in the spatial mode $M_i$, with the wave vector $k_{M_i} = k_W - k_{S_i}$, where $k_W$ and $k_{S_i}$ are the wave vectors of the write and Stokes fields, respectively. The atom-photon system can be written as [23, 29] $\rho_{ap}^{(ith)} = |0\rangle\langle 0| + \chi_i |\Phi\rangle^{(ith)(ith)}\langle\Phi|$, where $|\Phi\rangle^{(ith)} = \sqrt{\chi_i}\left(\cos\vartheta |+\rangle_{M_i} |R\rangle_{S_i} + \sin\vartheta |-\rangle_{M_i} |L\rangle_{S_i}\right)$ is the $i$th spin-wave-photon entangled state, $|+\rangle_{M_i}$ ($|-\rangle_{M_i}$) represents one spin-wave excitation associated with Zeeman coherence $|a, m_a\rangle \leftrightarrow |b, m_b = m_a\rangle$ ($|a, m_a\rangle \leftrightarrow |b, m_b = m_a + 2\rangle$), $|R\rangle_{S_i}$ ($|L\rangle_{S_i}$) denotes a $\sigma^+$ ($\sigma^-$)-polarized Stokes photon, $\cos\vartheta$ is the relevant Clebsch-Gordan coefficient [23]. In the presented experiment, we transform the $\sigma^+/\sigma^-$-polarization of the Stokes photon $S_i$ into the *H-V*-polarization by using a $\lambda/4$ plate. The Stokes photon $S_i$ is collected by a single-mode fiber ($SMF_{S_i}$), and then guided to a polarization-beam-splitter ($PBS_{S_i}$) which transmits horizontal (*H*) polarization into detector $D_{S_i}^{(1)}$ and reflects vertical (*V*) polarization into detector $D_{S_i}^{(2)}$. The detection events at the detectors $D_{S_i}$ ($i=1, ..., 6$) are processed by a Field Programmable Gate Array (FPGA) for further analysis. Once a photon is detected by the $i$th detector $D_{S_i}$ (either $D_{S_i}^{(1)}$ or $D_{S_i}^{(2)}$), one spin-wave excitation has been heralded to be stored in the mode $M_i$ and the subsequent write sequence is stopped by the feed-forward signal from the FPGA. After a storage time $\delta t$, a reading laser pulse is applied to convert the spin-wave excitation $M_i$ into the anti-Stokes photon $T_i$. By using a $\lambda/4$ plate, we transform the $\sigma^+/\sigma^-$-polarization of the photon $T_i$ into the *H-V*-polarization. So, the atom-photon state is transformed into the entangled two-photon state $|\Phi\rangle_{S,T}^{'(ith)} = \left(\cos\vartheta |H\rangle_{T_i}|H\rangle_{S_i} + \sin\vartheta |V\rangle_{T_i}|V\rangle_{S_i}\right)$. The photon $T_i$ passes through the single-mode fiber $\mathrm{SMF}_{T_i}$ and is sent to $i$th input port of an $m\times 1$ OSN. Based on the feed-forward-controlled signal, the OSN routes the $T_i$ photon into a common single-mode-fiber (CSMF). Passing through the CSMF, the photon $T_i$ impinges onto a polarization-beam-splitter (PBS$_T$), which transmits an *H*-polarized photon (reflects *V*-polarized photon) to detector $D_T^{(1)}$ ($D_T^{(2)}$).

The preparation rate of the atom-photon (photon-photon) entanglement pair can be evaluated by the rate of detected Stokes photons (coincidence count) in the *H-V* polarization setting. For the non-multiplexed case, in which only one source, for example, the $i$th source, is operating, the Stokes detection (coincidence count) rate for the source is given by $R_{H-V}^{(ith)} = r\left\{p_{SH}^{(ith)} + p_{SV}^{(ith)}\right\}$ ( $C_{H-V}^{(ith)} = C_{SH,TH}^{(ith)} + C_{SV,TV}^{(ith)}$ ), where, $r$ is repetition rate, $p_{SH}^{(ith)}$ ( $p_{SV}^{(ith)}$ ) is the probability of detecting a photon at the detector $D_{S_i}^{(1)}$ ( $D_{S_i}^{(2)}$ ), $C_{SH,TH}^{(ith)}$ ( $C_{SV,TV}^{(ith)}$ ) is the coincidence count rate between the detectors $D_{S_i}^{(1)}$ ( $D_{S_i}^{(2)}$ ) and $D_T^{(1)}$ ( $D_T^{(2)}$ ). For the multiplexed case, $m$ SWPE sources are simultaneously excited and the OSN is used for routing the anti-Stokes photons. In this case, the total Stokes detection (coincidence count) rate is written as $R_{H-V}^{(m)} = r\sum_{i=1}^{m}\left\{p_{SH}^{(M,ith)} + p_{SV}^{(M,ith)}\right\}$ ( $C_{H-V}^{(m)} = \sum_{i=1}^{m} C_{H-V}^{(M,ith)} = \sum_{i=1}^{m}\left\{C_{SH,TH}^{(M,ith)} + C_{SV,TV}^{(M,ith)}\right\}$ ), where, $p_{SH}^{(M,ith)}$ ( $p_{SV}^{(M,ith)}$ ) is the probability of detecting a photon at the detector $D_{S_i}^{(1)}$ ($D_{S_i}^{(2)}$), $C_{SH,TH}^{(M,ith)}$ ($C_{SV,TV}^{(M,ith)}$) is the coincidence count rate between the detectors $D_{S_i}^{(1)}$ ($D_{S_i}^{(2)}$) and $D_T^{(1)}$ ($D_T^{(2)}$), the superscript $M$ denotes the multiplexed case.

To verify that the MI can enhance the probability for generating entangled atom-photon (photon-photon) pairs while introducing no extra noise, we measure the polarization visibilities versus the Stokes detection (coincidence count) rate for the multiplexed and non-multiplexed case. For the non-multiplexed case, the polarization visibility of the two photons from the source $i$ in the *H-V* polarization setting is defined as $V_{H-V}^{(ith)} = (C_{H-V}^{(ith)} - N_{H-V}^{(ith)})/(C_{H-V}^{(ith)} + N_{H-V}^{(ith)})$, where, $C_{H-V}^{(ith)} = C_{SH,TH}^{(ith)} + C_{SV,TV}^{(ith)}$ ($N_{H-V}^{(ith)} = C_{SH,TV}^{(ith)} + C_{SV,TH}^{(ith)}$) denotes the coincidence count rate, which is measured when the router circuitry consisting of OSN and CSMF is removed. For the multiplexed case, the polarization visibility of the entangled two photons from the MI in *H-V* basis is defined as [see Eq. (S37) in Supplemental Material [56]] $V_{H-V}^{(M)} = \sum_{i=1}^{m=6} P_{S,T}^{(M,ith)} V_{H-V}^{(M,ith)} \approx \left( C_{H-V}^{(m)} - N_{H-V}^{(m)} \right) / \left( C_{H-V}^{(m)} + N_{H-V}^{(m)} \right)$ where, $N_{H-V}^{(m)} = \sum_{i=1}^{m} N_{H-V}^{(M,ith)} = \sum_{i=1}^{m} \left( C_{SH,TV}^{(M,ith)} + C_{SV,TH}^{(M,ith)} \right)$, $P_{S,T}^{(M,ith)} = \left( C_{H-V}^{(M,ith)} + N_{H-V}^{(M,ith)} \right) / \left( C_{H-V}^{(m)} + N_{H-V}^{(m)} \right)$ represents the normalized coincidence probability, $C_{SH,TV}^{(M,ith)}$ ($C_{SV,TH}^{(M,ith)}$) is the coincidence count rate between the detectors $D_{S_i}^{(1)}$ ($D_{S_i}^{(2)}$) and $D_T^{(2)}$ ($D_T^{(1)}$).

Before showing the visibilities as a function of the Stokes detection (coincidence count) rates, we measured the two-photon coincidences of the MI in *H-V*, $D-A = +/-45°$ and *R-L* polarization settings for a fixed value of $p_S^{(m=6)} \approx 1.26\%$, where, $p_S^{(m=6)} = \sum_{i=1}^{m=6} p_S^{(M,ith)}$ is the total Stokes-detection probability, with $p_S^{(M, ith)}$ being the probability detecting a photon at the detector $D_{S_i}$ ( either $D_{S_i}^{(1)}$ or $/D_{S_i}^{(2)}$ ) for the multiplexed case. The measured results are plotted in the histogram shown in Fig.3. From the results, we calculated the visibilities of the MI for *H-V, D-A* and *R-L* polarization settings according to the Eq. (S37) in Supplemental Material, which are $V_{H-V}^{(M)} \approx 94.1\%$ , $V_{D-A}^{(M)} \approx 84.4\%$ and $V_{R-L}^{(M)} \approx 81.6\%$ , respectively.

The red, yellow, pink, green, blue and purple data in the Fig. 4(a) [4(b)] are the measured polarization visibilities $V_{H-V}^{(1st)}$, $V_{H-V}^{(2nd)}$, …and $V_{H-V}^{(6th)}$ as the functions of the Stokes detection [coincidence count] rate in the *H-V* polarization setting for the individual sources 1, 2, …, 6, respectively, which are measured under the non-multiplexed case. The black data in Fig. 4(a) and 4(b) are the measured visibilities for the MI versus $R_{H-V}^{(m)}$ and $C_{H-V}^{(m)}$, respectively. The blue (red) solid lines in Fig. 4(a) and 4(b) are the linear fittings to the single-source data (multiplexed data $V_{H-V}^{(M)}$ ). In the measurements, we fix the storage time at $\delta t = 1\mu s$ and increase the Stokes-detection (coincidence count) rates by increasing the write-light power. As shown in Fig. 4(a) and 4(b), the polarization visibilities $V_{H-V}^{(ith)}$ ( $i = 1, 2, ..., 6$ ) and $V_{H-V}^{(M)}$ decrease with the increase in the Stokes-detection (coincidence count) rates due to multi-excitation noise. However, for a fixed visibility, the MI gives rise to a 5.94-fold (3.98-fold) increase in the Stokes detection rate (coincidence count rate) compared to the non-multiplexed interface. These results agree with the theoretical prediction of $m = 6$ [ $\bar{\eta}_{RC} \times m \approx 4.1$ ] based on Eq.(S57) [Eq.(S59)] in Supplemental Material [56], indicating that the application of the router circuitry controlled by the feed-forward control doesn't introduce extra noise. Similarly to the measurements in Fig. 4, we also measure the dependences of the visibilities on the Stokes detection (coincidence count) rates in $D-A$ and $R-L$ polarization settings. The measured results are shown in Fig. S1 and S2 in the Supplementary material [56], which show that the MI enables a ~5.9-fold (~3.9-fold) increase in the probability of generating entangled atom-photon (photon-photon) pairs without introducing extra noise.

The quality of the created entanglement in the MI can be characterized by the compositive fidelity which is defined by $F^{(M)} \approx \sum_{i=1}^{m=6} P_{S,T}^{(M,ith)} F^{(M,ith)}$ (see Eq. (S2) in Supplemental Material [56]), where $P_{S,T}^{(M,ith)} = p_{S,T}^{(M,ith)} / \sum_{i=1}^{m} p_{S,T}^{(M,ith)}$, $p_{S,T}^{(M,ith)}$ is the coincidence probability between the detectors $D_{S_i}$ (either $D_{S_i}^{(1)}$ or $D_{S_i}^{(2)}$) and $D_T$ (either $D_T^{(1)}$ or $D_T^{(2)}$), $F^{(M,ith)} = \left( Tr\sqrt{\sqrt{\rho_r^{(M,ith)}} \rho_d \sqrt{\rho_r^{(M,ith)}}} \right)^2$, $\rho_r^{(M,ith)}$ is the reconstructed density matrix of the $i$th entangled atom-photon state, $\rho_d$ is the density matrix of the entangled state defined by $|\Phi\rangle_{S,T}^{'(ith)}$. The probability $P_{S,T}^{(M,ith)}$ and fidelities $F^{(M,ith)}$ ($i=1$ to $6$) are directly measured under the multiplexed case. Fig.5 (a) and (b) plot the reconstructed density matrices $\rho_r^{(M,1st)}$ and $\rho_r^{(M,3th)}$, respectively, which yield $F^{(M,1st)} = 87.1\%$ and $F^{(M,3th)} = 88.0\%$. Table I shows the measured $F^{(M)}$ for several different values of $p_S^{(m=6)}$ for a delay time of 1μs. For $p_S^{(6)} \approx 1.26\%$ corresponding to the write-light power to obtain $C_{H-V}^{(6)} \approx 200 s^{-1}$ (see Fig. 4), we measure a maximal fidelity of $F^{(M)} = 87.9\%$.

The quality of the entangled atom-photon states created in the MI can also be described by a compositive Bell parameter $S^{(M)}$, which is defined as (see Eq. (S105) in Supplemental Material [56]),
$S^{(M)} = \sum_{i=1}^{m=6} P_{S,T}^{(M,ith)} S^{(M,ith)} = |E^{(M)}(\theta_{S_i},\theta_T) - E^{(M)}(\theta_{S_i},\theta_T') + E^{(M)}(\theta_{S_i}',\theta_T) + E^{(M)}(\theta_{S_i}',\theta_T')| < 2$, where, $S^{(M,ith)}$ is the Bell parameter between the photons $S_i$ and $T_i$, $\theta_{S_i}$ ($\theta_T$) is the polarization angle of the Stokes (anti-Stokes) field, which is set by rotating a $\lambda/2$ plate before $PBS_{S_i}$ ($PBS_T$), $E^{(M)}(\theta_{S_i},\theta_T)$ is the correlation function defined by:

$$\frac{\sum_{i=1}^{m} \left( C_{SH,TH}^{(M,ith)}(\theta_{S_i},\theta_T) + C_{SV,TV}^{(M,ith)}(\theta_{S_i},\theta_T) - C_{SH,TV}^{(M,ith)}(\theta_{S_i},\theta_T) - C_{SV,TH}^{(M,ith)}(\theta_{S_i},\theta_T) \right)}{\sum_{i=1}^{m} \left( C_{SH,TH}^{(M,ith)}(\theta_{S_i},\theta_T) + C_{SV,TV}^{(M,ith)}(\theta_{S_i},\theta_T) + C_{SH,TV}^{(M,ith)}(\theta_{S_i},\theta_T) + C_{SV,TH}^{(M,ith)}(\theta_{S_i},\theta_T) \right)}, \quad (1)$$

for example, $C_{SH,TH}^{(M,ith)}(\theta_{S_i},\theta_T)$ ($C_{SV,TV}^{(M,ith)}(\theta_{S_i},\theta_T)$) is the coincidence detection rate between the detectors $D_{S_i}^{(1)}$ ($D_{S_i}^{(2)}$) and $D_T^{(1)}$ ($D_T^{(2)}$) for the polarization angle $\theta_{S_i}$ and $\theta_T$. In the measurement, the canonical settings are chosen to be $\theta_{S_i} = 0°$, $\theta_{S_i}' = 45°$ ($i=1$ to $6$), $\theta_T = 22.5°$ and $\theta_T' = 67.5°$. Table I also shows the measured data of $S^{(M)}$ for several different values of $p_S^{(6)}$ when $\delta t = 1\mu s$, For $p_S^{(6)} \approx 0.0126$, we achieve a maximal value of $S^{(M)} = 2.49 \pm 0.03$, violating Bell-CHSH inequality by ~16 standard deviations.

To investigate the ability to use the atomic ensemble for quantum memory applications, we measure the decay of the Bell parameter $S^{(M)}$ with the storage time $\delta t$. In the measurements, the peak power of write pulse is fixed to get $p_S^{(6)} = 0.0297$ (corresponding to $C_{H-V}^{(6)} \approx 400 s^{-1}$). The blue square dots in Fig. 6 depict the measured $S^{(M)}$ data, which shows that even after storing the spin-wave excitation for 51 µs, violates the corresponding CHSH-inequality by 2.9 standard deviations ($S^{(M)} = 2.23 \pm 0.08$). The red circle dots are the average Bell parameter defined by $\bar{S} = \left( S^{(M,1st)} + S^{(M,2nd)} + ... + S^{(M,mth)} \right)/m$, which is consistent with the values of $S^{(M)}$, indicating that the six-SWPE sources are approximately symmetric in the detection and retrieval efficiency [see Eq. (S111) in Supplemental Material [56]].

In summary, a key advance of our demonstrated multiplexed LMEI is the enhanced rate for generating entangled atom-photon (photon-photon) pairs without introducing additional noise. Several criteria such as compositive visibility, fidelity, and Bell parameter are applied for judging the quality of the entanglement created in the MI. The transmission of OSN remains unchanged when the multi-mode number scales up. To apply such multiplexed LMEI into long-distance QR applications, its several quantities need to be further improved. The lower retrieval efficiency (~15%) can be increased by using the high optical-depth cold atoms or coupling the atoms into an optical cavity [21]. The short storage lifetime (~51µs) can be extended by trapping atoms in an optical lattice [21] and selecting two magnetic-field-insensitive spin waves to store memory qubits [57]. The low multimode number can be further extended by collecting Stokes photons at more directions [58, 59]. When the presented spatially-multiplexed scheme is combined with temporally-multiplexed storage approaches [60-63], one could achieve a multiplexing of a large number of modes, which will significantly improve entanglement distribution rates in long-distance quantum communications.

**Acknowledgments**

The authors are grateful to the referees for constructive and helpful suggestions. We also thank Professor Jing Zhang for helpful suggestions. We acknowledge funding support from Key Project of the Ministry of Science and Technology of China (2016YFA0301402), the National Natural Science Foundation of China (No.11475109, 11274211, 11604191), the Program for Sanjin Scholars of Shanxi Province of China.

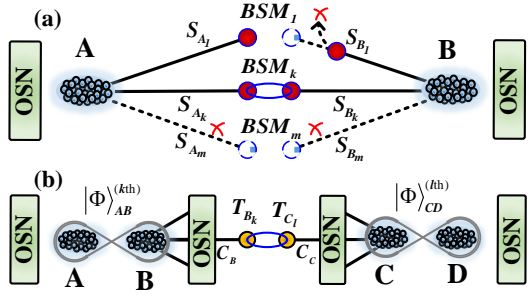

FIG. 1. (color online) Quantum repeater scheme based on the multiplexed interfaces. (a) Entanglement creation in an elementary link. A and B ensembles are located at left and right end points of the link, respectively. Each one is synchronically excited by write-laser pulses, which generates $m$ SWPE sources, labeled by $|\Phi\rangle_{A(B)}^{(1\text{st})},\ldots|\Phi\rangle_{A(B)}^{(i\text{th})},\ldots|\Phi\rangle_{A(B)}^{(m\text{th})}$. The SWPE source $|\Phi\rangle_A^{(i\text{th})}$ ($|\Phi\rangle_B^{(i\text{th})}$) emits a Stokes photon $S_{A_i}$ ($S_{B_i}$) and creates one spin-wave excitation $M_{A_i}$ ($M_{B_i}$). The photons $S_{A_i}$ and $S_{B_i}$ are sent to the $i$th middle station between A and B ensembles for BSM. Conditioned on a successful BSM at the $k$th station, for example, the ensembles A and B are projected into an entangled state $|\Phi\rangle_{AB}^{(k\text{th})}$. (b) Entanglement swapping between two adjacent links. The ensemble C (D) also creates $m$ SWPE sources and a successful BSM at the $l$th central station of the C-D link, e.g., projects the C and D ensembles into an entangled state $|\Phi\rangle_{CD}^{(l\text{th})}$. To performing an entanglement swapping between the A-B and C-D links, we convert the spin-wave excitation $M_{B_k}$ ($M_{C_l}$) into an anti-Stokes photon $T_{B_k}$ ($T_{C_l}$) and route it into a common channel $C_B$ ($C_C$) by using a feed-forward-controlled OSN.

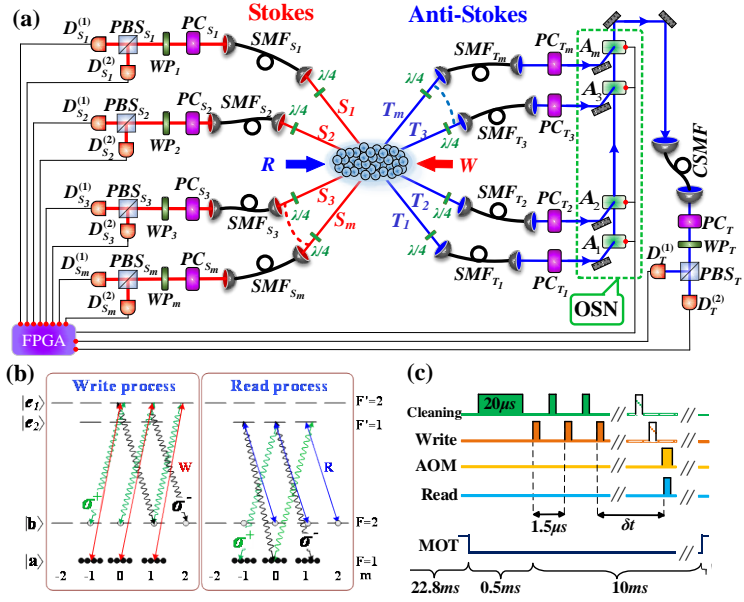

FIG. 2. Overview of the experiment. (a) Experiment setup for the 6-SWPE sources in combination with feed-forward control (Noted that we only plot 4-SWPE sources in the figure). PC: phase compensator (see Sec. II in Supplementary material [56]); $A_{(1,2...6)}$: acousto-optic modulators; $WP_{1,2,...6,T}$: half-wave/quarter-wave plates. In the measurements of the fidelities (Fig.5) or polarization visibilities (Fig.3), the $WP_{1,2,...6,T}$ are half-wave (quarter-wave) plates when analyzing the photon polarization in *D-A* (*R-L*) polarization setting and are removed when analyzing the photon polarization in *H-V* polarization setting [see Sec.V in Supplementary material [56]]. In the measurements of the Bell parameter, the $WP_{1,2,...6,T}$ are half-wave plates and used for setting the polarization angles. (b) Relevant atomic levels. (c) Time sequence of an experimental cycle.

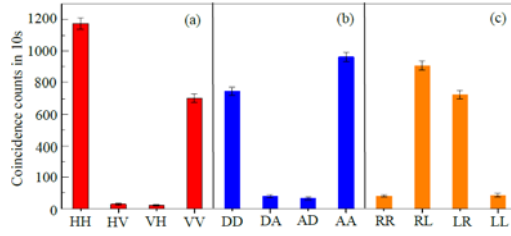

FIG. 3 (color online). Coincidences between the Stokes and anti-Stokes photons from the multiplexed LMEI at (a) *H-V* (b) *D-A* (c) *R-L* polarization settings for $p_S^{(m=6)} \approx 1.26\%$ and $\delta t = 1 \mu s$. Error bars represent $\pm 1$ standard deviation.

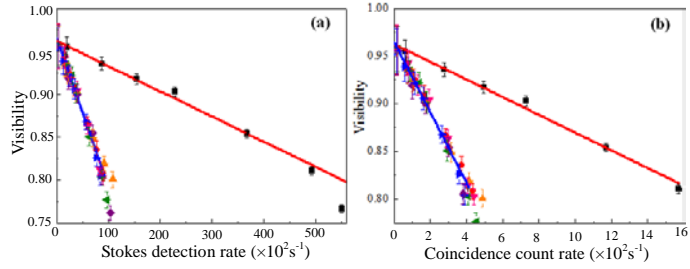

FIG. 4. The polarization visibility as functions of Stokes detection rate (a) and coincidence count rate (b) for the polarization setting of *H-V*, respectively. The blue solid lines in (a) and (b) are the least-square fittings to the single-source data according to the Eqs. (S48) and (S49) in Supplemental Material [56], respectively. While, the red solid lines in (a) and (b) are the least-square fittings to the multiplexed data $V_{H-V}^{(M)}$ according to the Eqs. (S57) and (S59) in Supplemental Material [56], respectively.

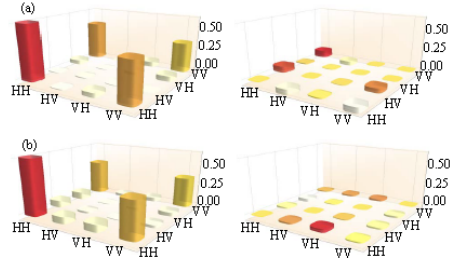

FIG. 5 (color online). Real and imaginary parts of the reconstructed density matrices $\rho_r^{(M,\,1st)}$ and $\rho_r^{(M,\,3th)}$ of the two photons from the sources 1 (a) and 3 (b), respectively.

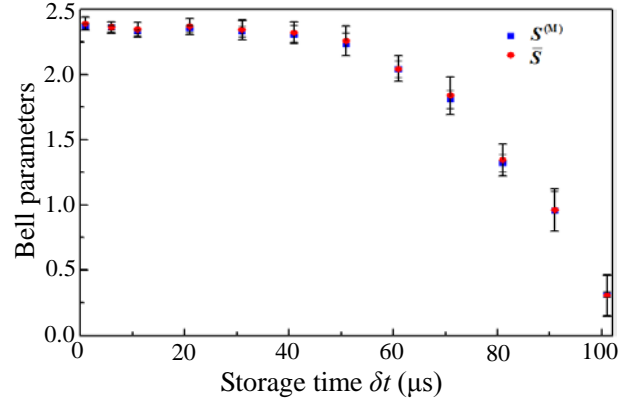

FIG. 6 (color online). Measurements of the Bell parameters $S^{(M)}$ and $\bar{S}$ as a function of $\delta t$ for $p_S^{(6)} = 0.0297$. Error bars represent $\pm 1$ standard deviation.

| $p_S^{(6)}$ | 0.0126 | 0.0297 | 0.0421 | 0.0594 | 0.0738 |
|---|---|---|---|---|---|
| $F^{(M)}$ | 0.87(1) | 0.85(1) | 0.82(1) | 0.78(1) | 0.75(1) |
| $S^{(M)}$ | 2.49(3) | 2.38(1) | 2.29(2) | 2.17(2) | 2.09(2) |

Table I. Measurement of the fidelity $F^{(M)}$ and Bell parameter $S^{(M)}$ for several different values of the total-Stokes-detection probability $p_S^{(6)}$ at $\delta t = 1\mu s$. Error bars represent $\pm 1$ standard deviation.

**Spatial multiplexing of atom-photon entanglement sources using feed-forward control and switching networks**
**Supplementary Material**

*Long Tian, Zhongxiao Xu, Lirong Chen, Wei Ge, Haoxiang Yuan, Yafei Wen, Shengzhi Wang, Shujing Li and Hai Wang\**

*The State Key Laboratory of Quantum Optics and Quantum Optics Devices,*

*Collaborative Innovation Center of Extreme Optics,*

*Institute of Opto-Electronics, Shanxi University, Taiyuan, 030006,*

*People's Republic of China*

## I. Experimental details

As shown in Fig. 2(a) in the main text, ~$10^9$ $^{87}$Rb atoms are trapped in a two-dimension magneto-optical trap (MOT) with a size of ~$5\times 2\times 2 mm^3$, a temperature of ~$130\mu K$ and an optical density of about 7. The write and read beams counterpropagate through the atoms, whose diameters (powers) in the MOT are ~3mm (100μW) and ~3.3mm (~50 mW), respectively. The writing light field (W) is tuned to the transition $|a\rangle \leftrightarrow |e_1\rangle$ with a 20MHz blue detuning. The reading light field (R) is on resonance with the transition from $|b\rangle \leftrightarrow |e_2\rangle$ [see Fig. 2(b) in the main text]. The atoms are optically pumped into the initial level $|a\rangle$ by a cleaning laser beam which is the combination of the laser beams $P_{CL1}$ and $P_{CL2}$ at a polarization-beam-splitter (PBS) and goes through the atoms at an angle of $2°$ to the z-axis. The $P_{CL1}$ and $P_{CL2}$ laser beams are $\sigma^{\pm}$-polarized, whose frequencies are tuned on the $|b\rangle \leftrightarrow |e_2\rangle$ and $|b\rangle \leftrightarrow |e_1\rangle$ transitions, respectively, and powers are kept at ~60mW.

The time sequence of the experimental procedure is shown in Fig. 2(c) in the main text, with a frequency of ~30 Hz. In each experimental period, a 23.3-ms preparation stage and a 10-ms experiment run are carried out. During the preparation stage, the atomic ensemble are trapped in the MOT for 22.8 ms and further cooled by a Sisyphus cooling for 0.5ms. Then the MOT (beams and magnetic field) is turned off. At the end of this stage, the cleaning laser beams are switched on for $20\mu s$ to pump the atoms into the level $|a\rangle$. After the preparation stage, the 10-ms experimental run containing many experimental cycles starts. Each cycle consists of a sequence of $N$ writing trials finally followed by a read pulse. Each writing trial is formed by a write laser pulse with a duration of $\Delta t_W = 80 ns$ and a cleaning pulse with a duration of $\Delta t_C = 250 ns$. The time interval between the two pulses is 670ns. The write pulse is applied on the atomic ensemble to generate correlated pairs of a single Stokes photon and a single SW excitation, while the cleaning pulse to pump the atoms into the initial level $|a\rangle$. The interval between two adjacent write trials is 1.5μs, corresponding to a repetition rate of $r \approx 6.7 \times 10^5 / s$. The detection events at the detectors $D_{S_1}$ ($D_{S_1}^{(1)}$ or $D_{S_1}^{(2)}$), $D_{S_2}$ ($D_{S_2}^{(1)}$ or $D_{S_2}^{(2)}$), …, $D_{S_6}$ ($D_{S_6}^{(1)}$ or $D_{S_6}^{(2)}$) are analyzed by a FPGA. As soon as a Stokes photon is detected by any one of these detectors, for example, the $S_i$ photon is detected by the detector $D_{S_1}$ (either $D_{S_1}^{(1)}$ or $D_{S_1}^{(2)}$), the FPGA will send out a feed-forward signal to stop the write trials. After a storage time, a read laser pulse is applied on the atoms to convert the spin-wave excitation into the anti-Stokes photon $T_i$. At the same time, the FPGA delivers a feed-forward signal to the control circuit to switch on the $i$th acousto-optic modulator (AOM) of the optical switching network (OSN) and then the photon $T_i$ is routed into the common single-mode fiber (CSMF). After 770ns time interval, the next experimental cycle starts.

We use the OSN to route the anti-Stokes photons $T_i$ ($i = 1, 2, …, 6$) into the CSMF. The combination of OSN and CSMF forms a router circuitry. The OSN consists of $m$ AOMs. As shown in Fig. 2(a) in the main text, the anti-Stokes photon $T_i$ is collected by a single-mode fiber ($SMF_{T_i}$) and its output is coupled to the input port of the $i$th AOM ($A_i$) and then is routed into the CSMF by switching on the $A_i$. The total transmission of the router circuitry for the $i$th anti-Stokes optical field mode is $\eta_{RC_i} = \eta_{OSN_i} \cdot \eta_{CF_i}$, including the OSN transmission $\eta_{OSN_i}$ and CSMF coupling efficiency $\eta_{CF_i}$, which shows that the designed OSN can keep a constant transmission loss for each anti-Stokes optical field when multi-mode number scales up. The measured transmissions are in the supplementary table S1, which shows an average transmission $\bar{\eta}_{RC} = \sum_{i=1}^{m} \eta_{RC_i} / m = 68.3\%$, with deviations $\delta\eta_{RC_i} = (\eta_{RC_i} - \bar{\eta}_{RC}) / \bar{\eta}_{RC}$ of less than 1.4%.

In the $i$th Stokes channel [as shown in the Fig. 2(a)], we use the single-mode fiber $SMF_{S_i}$ to collect the Stokes photon $S_i$. After the $SMF_{S_i}$, the Stokes photon $S_i$ goes through an optical filter (say $Filter_{S_i}$) for filtering the background noise in the Stokes field. Then, the photon $S_i$ is directed to the polarization-beam-splitter $PBS_{S_i}$. The two outputs of the Stokes fields from $PBS_{S_i}$ are coupled into two multi-mode fibers and then directly sent into the single-photon detectors $D_{S_i}^{(1)}$ and $D_{S_i}^{(2)}$, respectively. The total detection efficiency for detecting the $i$th Stokes photons at the detector $D_{S_i}$ ($D_{S_i}^{(1)}$ or $D_{S_i}^{(2)}$) is given by $\eta_{S_i} = \eta_{SMF_{S_i}} \cdot \eta_{Filter_{S_i}} \cdot \eta_{MMF_{S_i}} \cdot \eta_{SPD}$, which includes the coupling efficiency $\eta_{SMF_{S_i}}$ of the single-mode fiber $SMF_{S_i}$, the transmission $\eta_{Filter_{S_i}}$ of the optical filter $Filter_{S_i}$, the coupling efficiency $\eta_{MMF_{S_i}}$ of the multi-mode fibers, and the quantum efficiency $\eta_{SPD} \approx 0.5$ of the detector $D_{S_i}$ ($D_{S_i}^{(1)}$ or $D_{S_i}^{(2)}$). The measured efficiencies are shown in the supplementary table S2. Using these data, we evaluated the total detection efficiencies $\eta_{S_1}=0.29$, $\eta_{S_2}=0.29$, $\eta_{S_3}=0.29$, $\eta_{S_4}=0.30$, $\eta_{S_5}=0.30$ and $\eta_{S_6}=0.29$, and then obtained the average total detection efficiency $\bar{\eta}_S = \frac{1}{m}\sum_{i=1}^{m=6}\eta_{S_i} = 29\%$, with deviations $\delta\eta_{S_i} = (\eta_{S_i} - \bar{\eta}_S)/\bar{\eta}_S$ being less than 2.7% [2.7%].

In the $i$th anti-Stokes channel [as shown in the Fig. 2(a)], we use the single-mode fiber $SMF_{T_i}$ to collect the anti-Stokes photon $T_i$. After the $SMF_{T_i}$, the photon $T_i$ goes through an optical filter (say $Filter_{T_i}$), which can filter the background noise in the anti-Stokes field. Then, the photon $T_i$ is directed into the polarization-beam-splitter $PBS_T$. The two outputs of the anti-Stokes fields from the $PBS_T$ are coupled into two multi-mode fibers and then directly sent into the single-photon detectors $D_{T_i}^{(1)}$ and $D_{T_i}^{(2)}$, respectively. The total detection efficiency for detecting the $i$th anti-Stokes photons at the detector $D_T$ ($D_T^{(1)}/D_T^{(2)}$) is given by $\eta_{T_i} = \eta_{SMF_{T_i}} \cdot \eta_{Filter_{T_i}} \cdot \eta_{MMF_T} \cdot \eta_{SPD}$, which includes the coupling efficiency $\eta_{SMF_{T_i}}$ of the fiber $SMF_{T_i}$, the transmission $\eta_{Filter_T}$ of the optical filter $Filter_{T_i}$, the coupling efficiency $\eta_{MMF_{T_i}}$ of the multi-mode fiber to single-photon detector and the quantum efficiency $\eta_{SPD} \approx 0.5$ of the detector $D_T$. The measured efficiencies are shown in the supplementary table S3. Using these data, we evaluated the total detection efficiencies $\eta_{T_1}=0.30$, $\eta_{T_2}=0.29$, $\eta_{T_3}=0.29$, $\eta_{T_4}=0.30$, $\eta_{T_5}=0.28$ and $\eta_{T_6}=0.29$, and then obtained the average total detection efficiency $\bar{\eta}_T = \frac{1}{m}\sum_{i=1}^{m=6}\eta_{T_i} = 29.2\%$, with deviations [$\delta\eta_{T_i} = (\eta_{T_i} - \bar{\eta}_T)/\bar{\eta}_T$] being less than 2.7%.

We measured the retrieval efficiencies of the *m*-SWPE sources for the non-multiplexed case, i.e., the case that only one source, for example, the *i*th source, is considered and the router circuitry which consists of the OSN and CSMF is removed. The retrieval efficiency for the *i*th anti-Stokes optical field mode may be measured as $\gamma_i \approx p_{S,T}^{(ith)} / (\eta_{T_i} \cdot p_S^{(ith)})$ [1, 2], where $p_S^{(ith)}$ is the probability of detecting a Stokes photon at the detector $D_{S_i}$ (either $D_{S_i}^{(1)}$ or $D_{S_i}^{(2)}$), $p_{S,T}^{(ith)}$ is the coincidence probability between the Stokes detector $D_{S_i}$ (either $D_{S_i}^{(1)}$ or $D_{S_i}^{(2)}$) and the anti-Stokes detector $D_T$ (either $D_T^{(1)}$ or $D_T^{(2)}$). The retrieval efficiency $\gamma_i$ corresponds to the probability to find an anti-Stokes photon before the fiber $SMF_{T_i}$ conditioned on detecting a Stokes photon. Based on the above expression, we calculated the retrieval efficiencies by using the experimental data of the probabilities $p_S^{(ith)}$ and $p_{S,T}^{(ith)}$ for the storage time of $\delta t = 1\mu s$, as well as the total detection efficiencies $\eta_{T_i}$, which are $\gamma_1 = 15.6\%$, $\gamma_2 = 15.6\%$, $\gamma_3 = 16.0\%$, $\gamma_4 = 15.1\%$, $\gamma_5 = 15.8\%$ and $\gamma_6 = 15.8\%$, respectively. The average retrieval efficiency is $\bar{\gamma}_i = \frac{1}{m}\sum_{i=1}^{m}\gamma_i = 15.7\%$, with deviations $\delta\gamma_i = (\gamma_i - \bar{\gamma}_i)/\bar{\gamma}_i$ being less than 2.6%.

## II. Polarization compensation

When an optical field passes through an optical element such as a single-mode fiber (SMF) or an acousto-optic modulator (AOM), the fidelity of its polarization state will be degraded due to the phase-shift difference between *H* and *V* polarizations of the optical field producing in the optical element. For avoiding such degradation, we use a phase compensator (PC) to eliminate the phase-shift difference between the *H* and *V* polarizations. The phase compensator is a combination of a λ/4, a λ/2 and a λ/4 wave-plates, which can generate any unitary transformation. As shown in Fig. 2(a) in the main text, the phase compensator $PC_{S_i}$ ($PC_{T_i}$) is used to compensate for the phase-shift difference resulting from the $SMF_{T_i}$ and the *i*th AOM ($A_i$) in OSN, the phase compensator $PC_T$ is used to compensate for the phase-shift difference resulting from the common single-mode fiber (CSMF).

**III. Error bars**

All error bars in the experimental data represent ±1 standard deviation, which are estimated from Poissonian detection statistic using Monte Carlo simulation.

**IV. The QR protocol based on the multiplexed light-matter entanglement interfaces (LMEIs)**

The basic procedure of the QR protocol that uses the multiplexed LMEI as basic building blocks is shown in Fig.1 in the main text. We assume that we want to distribute entanglement over a total distance $L=2L_0$, with $L_0$ being the distance of an elementary link, e.g., A-B link or C-D link. We start with entanglement generation for the A-B link as shown in Fig. 1(a) in the main text. Two ensembles are located in A and B nodes, respectively, each one is synchronically excited by a sequence of write-laser pulses, which can generate *m* spin-wave-photon entangled (SWPE) sources, say $|\Phi\rangle_{A(B)}^{(1st)}, \cdots |\Phi\rangle_{A(B)}^{(ith)}, \cdots |\Phi\rangle_{A(B)}^{(mth)}$, respectively. The SWPE source $|\Phi\rangle_A^{(ith)}$ ($|\Phi\rangle_B^{(ith)}$) can create an entangled pair of a

Stokes photon and one spin-wave (collective) excitation with an excitation probability of $\chi_{A(B)_i} \ll 1$ per write-pulse, the Stokes photon is in optical spatial mode $S_{A_i}$ ($S_{B_i}$) and the spin-wave excitation is in memory mode $M_{A_i}$ ($M_{B_i}$). The Stokes photons $S_{A_i}$ and $S_{B_i}$ are sent to the *i*th middle station between A and B ensembles for a joint Bell-state measurement (BSM). Conditioned on a successful BSM, for example, at the *k*th station, the atomic ensembles A and B are projected into the polarization entangled state $|\Phi\rangle_{AB}^{(k\text{th})}$ and the further writing-pulse sequence is stopped by a feed-forward-controlled circuitry formed by a Field Programmable Gate Array (FPGA). The ensemble C (D) [see Fig. 1(b)] also can create *m* SWPE sources, labeled as $|\Phi\rangle_{C(D)}^{(1\text{st})}, \cdots |\Phi\rangle_{C(D)}^{(i\text{th})}, \cdots |\Phi\rangle_{C(D)}^{(m\text{th})}$, respectively. Similar to the entanglement generation in the AB link, the entangled state $|\Phi\rangle_{CD}^{(l\text{th})}$ will be conditionally established on a successful BSM at the *l*th station between the C and D ensembles. Via entanglement swapping between the A-B and C-D links, we can establish entanglement between A and D ensembles. To accomplish such entanglement extension, we convert the spin-wave excitation $M_{B_k}$ ($M_{C_l}$) into the anti-Stokes photon $T_{B_k}$ ($T_{C_l}$) and route it into the common channel B (C) by using an optical switching network [OSN]. A single-mode fiber B (C) may be used as the common channel $C_B$ ($C_C$). The OSN used for routing $T_{B_k}$ ($T_{C_l}$) photon and the common single-mode fiber [CSMF] B (C) form a router circuitry. The router circuitries are assumed to have the same transmission $\bar{\eta}_{RC}$ (<1) in the present scheme. Passing through the common single-mode fibers B and C, respectively, the photons $T_{B_k}$ and $T_{C_l}$ are combined on a polarization-beam-splitter for BSM. A successful BSM at the station AD projects the A and D ensembles into an entangled state $|\Phi\rangle_{AD}$, which means that first level entanglement swapping is achieved.

In the above QR scheme, the $m$ SWPE sources in combination with an optical switching network controlled by a feed-forward circuitry, form a multiplexed LMEI. Due to the uses of the multiplexed LMEIs, a successful BSM at any of the middle stations will project the A (C) and B (D) ensembles into an entangled state. In this case, the total success probability for entanglement generation for the single link per write-pulse is $p_{S_{L0}}^{(m)} = 1 - \left(1 - p_{S_{L0}}^{(1)}\right)^m$ [see Eq. (S71) for detail], where $p_{S_{L0}}^{(1)}$ is the success probability for entanglement creation in the single link using the non-multiplexed LMEIs (e.g., 1 SWPE source) as basic building blocks. For $m p_{S_{L0}}^{(1)} \ll 1$, we have $p_{S_{L0}}^{(m)} \approx m p_{S_{L0}}^{(1)}$, showing that the multiplexing scheme can increase the probability of generating entanglement in a single link by a factor of $m$. For the presented repeater scheme in which $n=1$ nest level is required, this increase in the probability will lead to a decrease in the total time needed for the entanglement distribution over the distance $L$ by a factor of $m \bar{\eta}_{RC}^2$ (see Sec. IX for details), with $\bar{\eta}_{RC} = \bar{\eta}_{OSN} * \bar{\eta}_{CF}$ the transmission of the router circuitry, including the average OSN transmission $\bar{\eta}_{OSN}$ and CSMF coupling efficiency $\bar{\eta}_{CF}$.

The quality of the atom-atom entangled state between the ensembles A and B can be characterized by the fidelity $F_{AB}^{(M)}$ or visibility $V_{AB}^{(M)}$, which are approximately expressed as (see Eqs. (S98) and (S93) for details)

$$F_{AB}^{(M)} \approx \left(1+(4F_A^{(M)}-1)(4F_B^{(M)}-1)/3\right)/4, \tag{S1a}$$

$$V_{AB}^{(M)} \approx V_A^{(M)} V_B^{(M)}, \tag{S1b}$$

where $F_A^{(M)}$ ($F_B^{(M)}$) and $V_A^{(M)}$ ($V_B^{(M)}$) are the compositive fidelity and visibility of the multiplexed interface A (B), respectively, which are expressed by (see Eqs. (S97) and (S83) for details)

$$F_A^{(M)} = \sum_{i=1}^{m} P_{S_A,T_A}^{(M,ith)} F_A^{(M,ith)}, \tag{S2a}$$

$$F_B^{(M)} = \sum_{i=1}^{m} P_{S_B,T_B}^{(M,ith)} F_B^{(M,ith)}, \tag{S2b}$$

$$V_A^{(M)} = \sum_{i=1}^{m} P_{S_A,T_A}^{(M,ith)} V_A^{(M,ith)}, \tag{S2c}$$

$$V_B^{(M)} = \sum_{i=1}^{m} P_{S_B,T_B}^{(M,ith)} V_B^{(M,ith)}, \tag{S2d}$$

where, $P_{S_A,T_A}^{(M,ith)} = p_{S_A,T_A}^{(M,ith)} / \sum_{i=1}^{m} p_{S_A,T_A}^{(M,ith)}$ ($P_{S_B,T_B}^{(M,ith)} = p_{S_B,T_B}^{(M,ith)} / \sum_{i=1}^{m} p_{S_B,T_B}^{(M,ith)}$) is the normalized coincidence probability, $p_{S_A,T_A}^{(M,ith)}$ ($p_{S_B,T_B}^{(M,ith)}$) is the coincidence probability of detecting a pair of photons, one is at the Stokes channel and the other is at the anti-Stokes channel of the source $i$ in the ensemble A (B) (see Sec. X for details), $F_A^{(M,ith)}$ ($F_B^{(M,ith)}$) and $V_A^{(M,ith)}$ ($V_B^{(M,ith)}$) are the fidelity and visibility of the entangled Stokes and anti-Stokes photons from the source $i$ in the ensemble A (B), the superscript $M$ of the physical quantities such as $p_{S_A,T_A}^{(M,ith)}$, $F_A^{(M,ith)}$, $V_A^{(M,ith)}$ etc. denotes that the measurement of the physical quantities are in the multiplexed case.

## V. The generation rates of the entangled atom-photon (photon-photon) pairs from non-multiplexed and multiplexed LMEIs

The heart of the LMEIs is the atom-photon entanglement which is created by the SWPE sources. We first discuss the SWPE source and then give the generation rates of the entangled atom-photon (photon-photon) pairs from the non-multiplexed and multiplexed LMEIs

### (1) The SWPE source

The multiplexed LMEI can generate $m$ SWPE sources in an atomic ensemble. The $i$th source emits a Stokes photon into the spatial mode $S_i$ and simultaneously creates one atomic spin-wave excitation in the memory mode $M_i$ [see Fig. 2(a) in the main text] with a small probability $\chi_i$. As mentioned in the main text, the state of the $i$th atom-photon system can be expressed as

$$|\varphi\rangle^{(i\text{th})} = |0\rangle + \sqrt{\chi_i}\left(\cos\vartheta|+\rangle_{M_i}|R\rangle_{S_i} + \sin\vartheta|-\rangle_{M_i}|L\rangle_{S_i}\right), \quad (S3)$$

where, $R(L)$ denotes right/left circular $\sigma^+(\sigma^-)$ polarization of the single Stokes photon, $|+\rangle$ and $|-\rangle$ represent two states of the spin-wave excitation, the non-vacuum part can be written as

$$|\Phi\rangle^{(i\text{th})} = \left(\cos\vartheta|+\rangle_{M_i}|R\rangle_{S_i} + \sin\vartheta|-\rangle_{M_i}|L\rangle_{S_i}\right), \quad (S4)$$

which represents the entangled state between the Stokes photon and the atomic spin-wave excitation, i.e., SWPE state.

For convenience, we transform the $\sigma^+(\sigma^-)$-polarization state of the Stokes photons $S_i$ into the $H(V)$-polarization state by placing a $\lambda/4$ plate before the single-mode fiber $\text{SMF}_{S_i}$ [see the Fig. 2(a)]. Thus the SWPE state is

rewritten as:

$$|\Phi\rangle^{(ith)} = \left(\cos\vartheta |+\rangle_{M_i} |H\rangle_{S_i} + \sin\vartheta |-\rangle_{M_i} |V\rangle_{S_i}\right). \quad (S5)$$

The spin-wave excitation $|+\rangle(|-\rangle)$ will be converted into $\sigma^+(\sigma^-)$-polarized anti-Stokes photon $T_i$ when a reading laser pulse is applied to retrieve the spin-wave excitation. In the present experiment, we also transform the $\sigma^+(\sigma^-)$-polarization state of the anti-Stokes photons $T_i$ into the $H(V)$-polarization state by placing a $\lambda/4$ plate before the single-mode fiber $SMF_{T_i}$ [see the Fig. 2(a)]. So, the spin-wave excitation $|+\rangle(|-\rangle)$ will be converted into the $H(V)$-polarized anti-Stokes photon $T_i$ in the present experimental set up and the state $|\Phi\rangle^{(ith)}$ of Eq. (S5) can be viewed as the form of the SWPE state in the $H-V$ polarization setting.

Furthermore, we can rewrite the atom-photon state $|\varphi\rangle^{(ith)}$ of Eq. (S3) as the form in the $H-V$ polarization setting, which is

$$|\varphi\rangle^{(ith)} = |0\rangle + \sqrt{\chi_i}\left(\cos\vartheta |+\rangle_{M_i} |H\rangle_{S_i} + \sin\vartheta |-\rangle_{M_i} |V\rangle_{S_i}\right). \quad (S6)$$

In the $R$-$L$ ($\sigma^+-\sigma^-$) and $D$-$A$ (+45°/-45°) polarization settings, the state of Eq. (S6) can be rewritten as the two forms

$$|\varphi\rangle^{(ith)} = |0\rangle + \frac{\sqrt{\chi_i}}{2}\left((\cos\vartheta - \sin\vartheta)|\psi^+\rangle_{R-L}^{(ith)} + (\cos\vartheta + \sin\vartheta)|\psi^-\rangle_{R-L}^{(ith)}\right), \quad (S7a)$$

$$|\varphi\rangle^{(ith)} = |0\rangle + \frac{\sqrt{\chi_i}}{2}\left((\cos\vartheta + \sin\vartheta)|\psi^+\rangle_{D-A}^{(ith)} + (\cos\vartheta - \sin\vartheta)|\psi^-\rangle_{D-A}^{(ith)}\right), \quad (S7b)$$

respectively, where $|\psi^+\rangle_{R-L}^{(ith)} = |\delta^+\rangle_{M_i}|R\rangle_{S_i} + |\delta^-\rangle_{M_i}|L\rangle_{S_i}$, $|\psi^-\rangle_{R-L}^{(ith)} = |\delta^+\rangle_{M_i}|L\rangle_{S_i} + |\delta^-\rangle_{M_i}|R\rangle_{S_i}$, $|\psi^+\rangle_{D-A}^{(ith)} = |\Delta^+\rangle^{(ith)}|D\rangle^{(ith)} + |\Delta^-\rangle^{(ith)}|A\rangle^{(ith)}$ and $|\psi^-\rangle_{D-A}^{(ith)} = |\Delta^+\rangle^{(ith)}|A\rangle^{(ith)} + |\Delta^-\rangle^{(ith)}|D\rangle^{(ith)}$ are maximal entangled states, $|\delta^\pm\rangle^{(ith)} = (|+\rangle^{(ith)} \pm i|-\rangle^{(ith)})/\sqrt{2}$ correspond to two spin-wave excitation states, which will be converted into the $\sigma^+/\sigma^-$-polarized anti-Stokes

photon $T_i$ in the present experimental setup, $\left|\Delta^{\pm}\right\rangle^{(ith)} = (\left|+\right\rangle^{(ith)} \pm \left|-\right\rangle^{(ith)})/\sqrt{2}$ also are two spin-wave excitation states, which will convert into the $D/A$-polarized anti-Stokes photon $T_i$ in the present experimental setup.

The anti-Stokes photon $T_i$ and the Stokes photon $S_i$ are a pair of polarization-entangled photons.

The rate of preparing the entangled atom-photon (photon-photon) pairs can be evaluated through the Stokes-photon detection (coincidence count) rate in the $X$-$Y$= $H$-$V$, $R$-$L$ or $D$-$A$ polarization settings. In the following, we discuss the Stokes-photon detection (coincidence count) rates for the non-multiplexed and multiplexed cases.

**(2) Non-multiplexed case**

For the non-multiplexed case, an individual source, for example, the $i$th source is operating and the router circuitry consisting of OSN and CSMF is removed. The Stokes-photon detection rate $R_{X-Y}^{(ith)}$ in the $X$-$Y$=$H$-$V$, $R$-$L$ or $D$-$A$ polarization setting for the $i$th source is defined by

$$R_{X-Y}^{(ith)} = R_{SX}^{(ith)} + R_{SY}^{(ith)} = r\left(p_{SX}^{(ith)} + p_{SY}^{(ith)}\right), \tag{S8}$$

where, $r$ is repetition rate, $R_{SX}^{(ith)} = rp_{SX}^{(ith)}$ ($R_{SY}^{(ith)} = rp_{SY}^{(ith)}$) is the Stokes-detection rate at $X$-polarization ($Y$-polarization), $p_{SX}^{(ith)}$ ($p_{SY}^{(ith)}$) is the probability of detecting a Stokes photon at the detector $D_{S_i}^{(1)}$ ($D_{S_i}^{(2)}$) set to measure the $X$-polarized ($Y$-polarized) photon. We use the following steps to set the polarization states of the photons entering the detectors. As shown in Fig. 2(a) in the main text, the Stokes photon $S_i$ is sent to the $\text{PBS}_{S_i}$ and then directed into the detector $D_{S_i}^{(1)}$ or $D_{S_i}^{(2)}$. If no wave-plate is placed before polarization-beam-splitter $PBS_{S_i}$, the $PBS_{S_i}$ will transmits *H*-polarization into the detector $D_{S_i}^{(1)}$ and reflects *V*-polarization into the detector $D_{S_i}^{(2)}$. If a λ/4 (λ/2) wave-plate [labeled as $WP_i$ in the Fig. 2(a)] is placed before the $PBS_{S_i}$ to transform the *R-L*-circular (*D-A*-linear) -polarization state into the *H-V*-polarization state, the $PBS_{S_i}$ will transmit *R*-circularly (*D*-linearly) -polarized $S_i$ photon into the $D_{S_i}^{(1)}$ detector and reflect the *L*-circularly (*A*-linearly) –polarized $S_i$ photon into the $D_{S_i}^{(2)}$ detector.

The Stokes-photon detection rates in the three polarization settings should be the same and then we can write the following equations

$$R_{H-V}^{(ith)} = R_{R-L}^{(ith)} = R_{D-A}^{(ith)} = R^{(ith)} = rp_S^{(ith)}, \tag{S9}$$

where, $p_S^{(ith)}$ is the probability of detecting a Stokes photon at $D_{S_i}$ ($D_{S_i}^{(1)}$ or $D_{S_i}^{(2)}$), which don't depend on the polarization settings.

The above idea can be explained by the following discussion. The relations between the Stokes probabilities and the excitation $\chi_i$ can be obtained from the Eqs. (S6) and (S7). Considering imperfect detection efficiencies, we have

$$p_{SH}^{(ith)} = \chi_i \cos^2 \vartheta \eta_{S_i}, \tag{S10a}$$

$$p_{SV}^{(ith)} = \chi_i \sin^2 \vartheta \eta_{S_i}, \tag{S10b}$$

$$p_{SR}^{(ith)} = p_{SD}^{(ith)} = \chi_i \eta_{S_i} / 2, \tag{S10c}$$

$$p_{SL}^{(ith)} = p_{SA}^{(ith)} = \chi_i \eta_{S_i} / 2, \tag{S10d}$$

where, $\eta_{S_i}$ is the total detection efficiency in the *i*th Stokes channels. Based on the Eq. (S10), we obtain

$$R_{H-V}^{(ith)} = R_{R-L}^{(ith)} = R_{D-A}^{(ith)} = R^{(ith)} = rp_S^{(ith)} = r\chi_i\eta_{S_i}. \qquad (S11)$$

The atom-photon entangled pair can be transferred into the photon-photon entangled pair by releasing the stored spin-wave excitation $M_i$ into the anti-Stokes photon $T_i$. Passing through the single-mode fiber $SMF_{T_i}$ and the phase compensator $PC_{T_i}$, the anti-Stokes photon $T_i$ is sent into the $PBS_T$, which transmits the *H*-polarized photon into the detector $D_T^{(1)}$ and reflects *V*-polarized photon into the detector $D_T^{(2)}$. If a λ/4 (λ/2) wave-plate [labeled as $WP_T$ in the Fig. 2(a)] is placed before the $PBS_T$, the $\sigma^+(\sigma^-)$-circular [*D(A)*-linear] polarization state of the photon $T_i$ will be transformed into *H(V)*-polarization state. In this case, the $PBS_T$ will transmit the $\sigma^+$-circularly (*D*-linearly) –polarized photon into the detector $D_T^{(1)}$ while reflect the $\sigma^-$ – circularly (*A*-linearly) -polarized photon into the detector $D_T^{(2)}$.

The generation rate of photon-photon pairs from the *i*th source can be evaluated by coincidence count rate in the polarization setting *X-Y=H-V, D-A* or *R-L*. For *X-Y=H-V* or *D-A* polarization setting, the coincidence count rate is defined as

$$C_{X-Y}^{(ith)} = C_{SX,TX}^{(ith)} + C_{SY,TY}^{(ith)} = r\left(p_{SX,TX}^{(ith)} + p_{SY,TY}^{(ith)}\right), \qquad (S12)$$

where, $C_{SX,TX}^{(ith)} = rp_{SX,TX}^{(ith)}$ ($C_{SY,TY}^{(ith)} = rp_{SY,TY}^{(ith)}$), $p_{SX,TX}^{(ith)}$ ($p_{SY,TY}^{(ith)}$) is the coincidence probability between the detectors $D_{S_i}^{(1)}$ ($D_{S_i}^{(2)}$) and $D_T^{(1)}$ ($D_T^{(2)}$) both set to measure X (Y) –polarized photon. While for the *X-Y=R-L* polarization setting, the $C_{X-Y}^{(ith)}$ is defined as

$$C_{X-Y}^{(ith)} = C_{R-L}^{(ith)} = C_{SR,TL}^{(ith)} + C_{SL,TR}^{(ith)} = r\left(p_{SR,TL}^{(ith)} + p_{SL,TR}^{(ith)}\right), \qquad (S13)$$

where, $p_{SR,TL}^{(ith)}$ ($p_{SL,TR}^{(ith)}$) is the coincidence probability between the detectors $D_{S_i}^{(1)}$ ($D_{S_i}^{(2)}$) set to measure $\sigma^+$ ($\sigma^-$) –polarized photon and $D_T^{(2)}$ ($D_T^{(1)}$) set to measure $\sigma^-$ ($\sigma^+$) –polarized photon.

**(3) For the multiplexed case**

For the multiplexed case, *m* SWPE sources are simultaneously excited and the OSN is used for routing the anti-Stokes photons $T_i$ ($i=1$ to $m$) into the CSMF. After the CSMF, the anti-Stokes photons $T_i$ is directed into the anti-Stokes detector $D_T$.

The preparation rate of the atom-photon pairs from the multiplexed LMEI can be evaluated by the total Stokes-photon detection rate, i.e., the sum of the Stokes detection rates for the *m* SWPE sources, which is defined by

$$R_{X-Y}^{(m)} = \sum_{i=1}^{m} R_{X-Y}^{(M,ith)} = \sum_{i=1}^{m} \left( R_{SX}^{(M,ith)} + R_{SY}^{(M,ith)} \right) = r \sum_{i=1}^{m} \left( p_{SX}^{(M,ith)} + p_{SY}^{(M,ith)} \right), \quad (S14)$$

where, *X-Y=H-V*, *R-L* or *D-A* polarization setting, $R_{X-Y}^{(M,ith)}$ represent the Stokes detection rates for the source *i*, $p_{SX}^{(M,ith)}$ ($p_{SY}^{(M,ith)}$) is the probability of detecting a Stokes photon at the detector $D_{S_i}^{(1)}$ ($D_{S_i}^{(2)}$) which is set to measure *X* (*Y*) –polarized photon, the superscript *M* denotes that the measurements for the rates or probabilities are performed for the multiplexed case.

The relations between the Stokes detection probabilities for the multiplexed and non-multiplexed cases can be written as

$$p_{SX}^{(M,ith)} = \left(1 - \bar{p}_S\right)^{i-1} p_{SX}^{(ith)}, \quad (S15a)$$

$$p_{SY}^{(M,ith)} = \left(1 - \bar{p}_S\right)^{i-1} p_{SY}^{(ith)}, \quad (S15b)$$

where, $\bar{p}_S = \left(\sum_{i=1}^{m} p_S^{(ith)}\right)/m$ is the average Stokes detection probability for the *m*

SWPE sources. Based on the Eq. (S15), we obtain

$$R_{X-Y}^{(M,ith)} = (1-\bar{p}_S)^{i-1} p_S^{(ith)}. \tag{S16}$$

According to the above expression, we know that the Stokes-photon detection rates $R_{X-Y}^{(M,ith)}$ also doesn't depend on the polarization setting, i.e. we can write the following equations

$$R_{H-V}^{(M,ith)} = R_{R-L}^{(M,ith)} = R_{D-A}^{(M,ith)} = R^{(M,ith)} = rp_S^{(M,ith)}, \tag{S17}$$

where, $p_S^{(M,ith)} = (1-\bar{p}_S)^{i-1} p_S^{(ith)}$ is the probability of detecting a Stokes photon at the detector $D_{S_i}$ (either $D_{S_i}^{(1)}$ or $D_{S_i}^{(2)}$) for the multiplexed case. So, the total Stokes-detection rate $R_{X-Y}^{(m)}$ can be rewritten as

$$R_{X-Y}^{(m)} = \sum_{i=1}^{m} R^{(M,ith)} = \sum_{i=1}^{m} rp_S^{(M,ith)} = rp_S^{(m)}, \tag{S18}$$

where, $p_S^{(m)}$ is the total Stokes-detection probability for the multiplexed interface, which is defined by.

$$p_S^{(m)} = \sum_{i=1}^{m} p_S^{(M,ith)}. \tag{S19}$$

According to the Eqs. (S15 and S11), we have

$$R_{X-Y}^{(m)} = R^{(m)} = \sum_{i=1}^{m} rp_S^{(M,ith)} = r\sum_{i=1}^{m} (1-\bar{p}_S)^{i-1} p_S^{(ith)} = r\sum_{i=1}^{m} (1-\bar{p}_S)^{i-1} \chi_i \eta_{S_i}. \tag{S20}$$

Defining the average Stokes detection efficiency as $\bar{\eta}_S = \left(\sum_{i=1}^{m} \eta_{S_i}\right)/m$ and assuming that the deviation $\delta\eta_{S_i} = \bar{\eta}_S - \eta_{S_i}$ is very small, we can rewrite the total Stokes-photon detection rate as

$$R_{X-Y}^{(m)} \approx r\bar{\eta}_S \sum_{i=1}^{m} (1-\bar{p}_S)^{i-1} \chi_i = r\bar{\eta}_S \chi^{(m)}, \tag{S21}$$

where, $\chi^{(m)} = \sum_{i=1}^{m}(1-\bar{p}_S)^{i-1} \chi_i$ can be viewed as the total excitation probability of the multiplexed interface. For the present experiment, $(1-\bar{p}_S)^{i-1} \approx 1$, so we have

$$p_{SX}^{(M,ith)} \approx p_{SX}^{(ith)}, \tag{S22a}$$

$$p_{SY}^{(M,ith)} \approx p_{SY}^{(ith)}, \qquad (S22b)$$

$$p_{S}^{(M,ith)} \approx p_{S}^{(ith)}. \qquad (S22c)$$

In this case, we obtain

$$R_{X-Y}^{(m)} = R^{(m)} = \sum_{j=1}^{m} r p_S^{(M,ith)} \approx r \sum_{i=1}^{m} p_S^{(ith)} = r \sum_{i=1}^{m} \chi_i \eta_{S_i}. \qquad (S23)$$

The preparation rate of the photon-photon pairs from the multiplexed LMEI can be evaluated by the total coincidence counts measured in the polarization setting $X$-$Y$=$H$-$V$, $D$-$A$ or $R$-$L$. For $X$-$Y$=$H$-$V$ or $D$-$A$ polarization setting, the total coincidence count rate is defined as

$$C_{X-Y}^{(m)} = \sum_{i=1}^{m} C_{H-V}^{(M,ith)} = \sum_{i=1}^{m} \left( C_{SX,TX}^{(M,ith)} + C_{SY,TY}^{(M,ith)} \right), \qquad (S24)$$

where, $C_{H-V}^{(M,ith)} = C_{SX,TX}^{(M,ith)} + C_{SY,TY}^{(M,ith)}$, $C_{SX,TX}^{(M,ith)} = r p_{SX,TX}^{(M,ith)}$ and $C_{SY,TY}^{(M,ith)} = r p_{SY,TY}^{(M,ith)}$, $p_{SX,TX}^{(M,ith)}$ ($p_{SY,TY}^{(M,ith)}$) is the coincidence probability between the detectors $D_{S_i}^{(1)}$ ($D_{S_i}^{(2)}$) and $D_T^{(1)}$ ($D_T^{(2)}$) both set to measure $X$ ($Y$) –polarized photon and is measured under the multiplexed case. While for $X$-$Y$=$R$-$L$ polarization setting, the total coincidence count rate is defined as

$$C_{X-Y}^{(m)} = C_{R-L}^{(m)} = \sum_{i=1}^{m} C_{R-L}^{(M,ith)} = \sum_{i=1}^{m} \left( C_{SR,TL}^{(M,ith)} + C_{SL,TR}^{(M,ith)} \right), \qquad (S25)$$

where, $C_{R-L}^{(M,ith)} = C_{SR,TL}^{(M,ith)} + C_{SL,TR}^{(M,ith)}$, $C_{SR,TL}^{(M,ith)} = r p_{SR,TL}^{(M,ith)}$ and $C_{SL,TR}^{(M,ith)} = r p_{SL,TR}^{(M,ith)}$, $p_{SR,TL}^{(M,ith)}$ ($p_{SL,TR}^{(M,ith)}$) is the coincidence probability between the detectors $D_{S_i}^{(1)}$ ($D_{S_i}^{(2)}$) set to measure $\sigma^+$ ($\sigma^-$) –polarized photon and $D_T^{(2)}$ ($D_T^{(1)}$) set to measure $\sigma^-$ ($\sigma^+$) –polarized photon. The setting of the polarization before the detectors $D_{S_i}$ and $D_T$ for the multiplexed case is the same as that for the non-multiplexed case mentioned above.

The relations between the coincidence probabilities for the multiplexed and non-multiplexed cases can be expressed as

$$p_{SX,TX}^{(M,ith)} = (1-\bar{p}_S)^{i-1} \eta_{RC_i} p_{SX,TX}^{(ith)}, \tag{S26a}$$

$$p_{SY,TY}^{(M,ith)} = (1-\bar{p}_S)^{i-1} \eta_{RC_i} p_{SY,TY}^{(ith)}, \tag{S26b}$$

$$p_{SX,TY}^{(M,ith)} = (1-\bar{p}_S)^{i-1} \eta_{RC_i} p_{SX,TY}^{(ith)}, \tag{S26c}$$

$$p_{SY,TX}^{(M,ith)} = (1-\bar{p}_S)^{i-1} \eta_{RC_i} p_{SY,TX}^{(ith)}, \tag{S26d}$$

where $\eta_{RC_i}$ is the total transmission of the router circuitry for the $i$th anti-Stokes optical field mode. For the present experiment, $(1-\bar{p}_S)^{i-1} \approx 1$, we have

$$p_{SX,TX}^{(M,ith)} \approx \eta_{RC_i} p_{SX,TX}^{(ith)}, \tag{S27a}$$

$$p_{SY,TY}^{(M,ith)} \approx \eta_{RC_i} p_{SY,TY}^{(ith)}, \tag{S27b}$$

$$p_{SX,TY}^{(M,ith)} \approx \eta_{RC_i} p_{SX,TY}^{(ith)}, \tag{S27c}$$

$$p_{SY,TX}^{(M,ith)} \approx \eta_{RC_i} p_{SY,TX}^{(ith)}. \tag{S27d}$$

And then, we can obtain

$$C_{X-Y}^{(M,ith)} \approx \eta_{RC_i} C_{X-Y}^{(ith)}. \tag{S28}$$

## VI. The polarization visibility of the entangled two-photon state for the Non-multiplexed and multiplexed LMEIs

The polarization visibility for an entangled two-photon state can be evaluated by means of the measured values of the Bell parameter or the fidelity of the state, it can also be directly measured by observing the polarization correlations between the two photons in the *X-Y=H-V, D-A* or *R-L* polarization setting.

**(1) Non-multiplexed case**

For the non-multiplexed interface, an individual source, for example, the source $i$, is considered, the polarization visibility of the entangled two photons from the source $i$ is defined as

$$V_{X-Y}^{(ith)} = \frac{C_{X-Y}^{(ith)} - N_{X-Y}^{(ith)}}{C_{X-Y}^{(ith)} + N_{X-Y}^{(ith)}}, \tag{S29}$$

where, for $X\text{-}Y = H\text{-}V$ or $D\text{-}A$, $N_{X-Y}^{(ith)} = C_{SX,TY}^{(ith)} + C_{SY,TX}^{(ith)} = r\left(p_{SX,TY}^{(ith)} + p_{SY,TX}^{(ith)}\right)$ denotes the coincidence count rate for perpendicular polarizations, $p_{SX,TY}^{(ith)}$ ($p_{SY,TX}^{(ith)}$) is the coincidence probability between the detector $D_{S_i}^{(1)}$ ($D_{S_i}^{(2)}$) set to measure $X$ ($Y$) –polarized photon and $D_T^{(2)}$ ($D_T^{(1)}$) set to measure $Y$ ($X$) –polarized photon. While, for $X\text{-}Y = R\text{-}L$ polarization setting, $N_{X-Y}^{(ith)} = N_{R-L}^{(ith)} = C_{SR,TR}^{(ith)} + C_{SL,TL}^{(ith)} = r\left(p_{SR,TR}^{(ith)} + p_{SL,TL}^{(ith)}\right)$ denotes the coincidence count rate for parallel polarizations, $p_{SR,TR}^{(ith)}$ ($p_{SL,TL}^{(ith)}$) is the coincidence probability between the detector $D_{S_i}^{(1)}$ ($D_{S_i}^{(2)}$) and $D_T^{(1)}$ ($D_T^{(2)}$) both set to measure $\sigma^+$ ($\sigma^-$) –polarized photon.

The Eq. (S29) can be rewritten as:

$$V_{X-Y}^{(ith)} = \frac{\left|\left(p_{SX,TX}^{(ith)} + p_{SY,TY}^{(ith)}\right) - \left(p_{SX,TY}^{(ith)} + p_{SY,TX}^{(ith)}\right)\right|}{\left(p_{SX,TX}^{(ith)} + p_{SY,TY}^{(ith)}\right) + \left(p_{SX,TY}^{(ith)} + p_{SY,TX}^{(ith)}\right)}, \tag{S30}$$

which shows that the polarization visibility $V_{X-Y}^{(ith)}$ can be measured by observing the coincidence probabilities $p_{SX,TX}^{(ith)}$, $p_{SY,TY}^{(ith)}$, $p_{SX,TY}^{(ith)}$ and $p_{SY,TX}^{(ith)}$.

The coincidence probability between the Stokes detector $D_{S_i}$ ($D_{S_i}^{(1)}$ or $D_{S_i}^{(2)}$) and the anti-Stokes detector $D_T$ ($D_T^{(2)}$ or $D_T^{(1)}$) is defined by

$$p_{S,T}^{(ith)} = \left(p_{SX,TX}^{(ith)} + p_{SY,TY}^{(ith)}\right) + \left(p_{SX,TY}^{(ith)} + p_{SY,TX}^{(ith)}\right) = \left(C_{X/Y}^{(ith)} + N_{X/Y}^{(ith)}\right)/r, \tag{S31}$$

which should independent on the polarization settings. This view point can be explained by the following discussion.

According to the Eqs. (S6) and (S7), we may express the coincidence probabilities as:

$$p_{SH,TH}^{(ith)} = \chi_i \cos^2 \vartheta \, \gamma_i \eta_{S_i} \eta_{T_i},\tag{S32a}$$

$$p_{SV,TV}^{(ith)} = \chi_i \sin^2 \vartheta \, \gamma_i \eta_{S_i} \eta_{T_i},\tag{S32b}$$

$$p_{SH,TV}^{(ith)} = 0,\tag{S32c}$$

$$p_{SV,TH}^{(ith)} = 0,\tag{S32d}$$

$$p_{SR,TL}^{(ith)} = p_{SL,TR}^{(ith)} = (\cos\vartheta + \sin\vartheta)^2 \chi_i \gamma_i \eta_{S_i} \eta_{T_i}/4,\tag{S32e}$$

$$p_{SR,TR}^{(ith)} = p_{SL,TL}^{(ith)} = (\cos\vartheta - \sin\vartheta)^2 \chi_i \gamma_i \eta_{S_i} \eta_{T_i}/4,\tag{S32f}$$

$$p_{SD,TD}^{(ith)} = p_{SA,TA}^{(ith)} = (\cos\vartheta + \sin\vartheta)^2 \chi_i \gamma_i \eta_{S_i} \eta_{T_i}/4,\tag{S32g}$$

$$p_{SD,TD}^{(ith)} = p_{SA,TA}^{(ith)} = (\cos\vartheta + \sin\vartheta)^2 \chi_i \gamma_i \eta_{S_i} \eta_{T_i}/4,\tag{S32h}$$

where, $\eta_{T_i}$ is the detection efficiency in the anti-Stokes channel of the source $i$, $\gamma_i$ is the retrieval efficiency of the source $i$. So, we have

$$\begin{aligned}\left(p_{SH,TH}^{(ith)} + p_{SV,TV}^{(ith)}\right) + \left(p_{SH,TV}^{(ith)} + p_{SV,TH}^{(ith)}\right) &= \left(p_{SD,TD}^{(ith)} + p_{SA,TA}^{(ith)}\right) + \left(p_{SH,TV}^{(ith)} + p_{SV,TH}^{(ith)}\right) \\ &= \left(p_{SR,TR}^{(ith)} + p_{SL,TL}^{(ith)}\right) + \left(p_{SH,TV}^{(ith)} + p_{SV,TH}^{(ith)}\right) = \chi_i \gamma_i \eta_{S_i} \eta_{T_i}\end{aligned}.\tag{S33}$$

**(2) Multiplexed case**

For the multiplexed case, the retrieved anti-Stokes photons from any of the sources will be routed into the CSMF by the OSN. The polarization visibility of the two photons from the source $i$ is defined as

$$V_{X-Y}^{(M,ith)} = \frac{\left|C_{X-Y}^{(M,ith)} - N_{X-Y}^{(M,ith)}\right|}{C_{X-Y}^{(M,ith)} + N_{X-Y}^{(M,ith)}},\tag{S34}$$

where, for $X$-$Y$=$H$-$V$ or $D$-$A$ polarization setting, $N_{X-Y}^{(M,ith)} = C_{SX,TY}^{(M,ith)} + C_{SY,TX}^{(M,ith)}$, $C_{SX,TY}^{(M,ith)} = r p_{SX,TY}^{(M,ith)}$ and $C_{SY,TX}^{(M,ith)} = r p_{SY,TX}^{(M,ith)}$, for example, $p_{SX,TY}^{(M,ith)}$ is the coincidence probability between the detector $D_{S_i}^{(1)}$ (set to measure $X$–polarized photon) and $D_T^{(2)}$ (set to measure *Y*–polarized photon), *M* represents that the measurement is performed under the multiplexed case. While for *X-Y=R-L* polarization setting, $N_{X-Y}^{(M,ith)} = N_{R-L}^{(M,ith)} = C_{SR,TR}^{(M,ith)} + C_{SL,TL}^{(M,ith)} = r\left(p_{SR,TR}^{(M,ith)} + p_{SL,TL}^{(M,ith)}\right)$, $p_{SR,TR}^{(M,ith)}$ ($p_{SL,TL}^{(M,ith)}$) is the coincidence probability between the detector $D_{S_i}^{(1)}$ ($D_{S_i}^{(2)}$) and $D_T^{(1)}$ ($D_T^{(2)}$) both set to measure $\sigma^+$ ($\sigma^-$) –polarized photon.

For the multiplexed case, we define the coincidence probability between the Stokes detector $D_{S_i}$ ($D_{S_i}^{(1)}$ or $D_{S_i}^{(2)}$) and the anti-Stokes detector $D_T$ ($D_T^{(2)}$ or $D_T^{(1)}$) as

$$p_{S,T}^{(M,ith)} = \left(p_{SX,TX}^{(M,ith)} + p_{SY,TY}^{(M,ith)}\right) + \left(p_{SX,TY}^{(M,ith)} + p_{SY,TX}^{(M,ith)}\right) = \left(C_{X/Y}^{(M,ith)} + N_{X/Y}^{(M,ith)}\right)/r, \quad (S35)$$

which also shouldn't depend on the polarization settings. This view point can be explained according to the following theoretical analysis. Based on the Eq. (27), we may obtain

$$p_{S,T}^{(M,ith)} = \eta_{RC_i} p_{S,T}^{(ith)} \approx \eta_{RC_i} \chi_i \gamma_i \eta_{S_i} \eta_{T_i}. \quad (S36)$$

Since the parameters $\eta_{RC_i}$, $\chi_i$, $\gamma_i$, $\eta_{S_i}$ and $\eta_{T_i}$ are unrelated to polarizations in the presented scheme, $p_{S,T}^{(M,ith)}$ is independent on the polarization settings.

The quality of the created entanglement in the MI can be described by the compositive polarization visibility of the photon-photon entangled state generated from the MI [see Eq. (S2) or Eq. (S83) for details]. In *X-Y* (*=H-V*, *D-A* or *R-L*) polarization setting, it can be expressed as

$$V_{X-Y}^{(M)} = \sum_{i=1}^{m} P_{S,T}^{(M,ith)} V_{X/Y}^{(M,ith)}. \quad (S37)$$

where, $P_{S,T}^{(M,ith)} = p_{S,T}^{(M,ith)} / \sum_{i=1}^{m} p_{S,T}^{(M,ith)} = \left(C_{H-V}^{(M,ith)} + N_{H-V}^{(M,ith)}\right)/\left(C_{H-V}^{(m)} + N_{H-V}^{(m)}\right)$ is the normalized coincidence probability. According to the expression of Eq. (S37) and the definition $V_{X-Y}^{(\text{M,ith})}$ of Eq. (S34), we obtain

$$V_{X-Y}^{(\text{M})} = \frac{C_{X-Y}^{(m)} - N_{X-Y}^{(m)}}{C_{X-Y}^{(m)} + N_{X-Y}^{(m)}}, \tag{S38}$$

where $N_{X-Y}^{(m)} = \sum_{i=1}^{m}\left(C_{SX,TY}^{(\text{M,ith})} + C_{SY,TX}^{(\text{M,ith})}\right)$, $C_{SX,TY}^{(\text{M,ith})} = rp_{SX,TY}^{(\text{M,ith})}$ and $C_{SY,TX}^{(\text{M,ith})} = rp_{SY,TX}^{(\text{M,ith})}$.

## VII. The relation between the polarization visibilities and excitation probability

We now develop a theory to model the relations between the polarization visibilities and the excitation probabilities for the multiplexed and non-multiplexed interfaces.

**(1) Non-multiplexed case**

In the above Sec.VI, we give the definitions of the Stokes detection probabilities $p_{SX}^{(\text{ith})}$ and $p_{SY}^{(\text{ith})}$, and Stokes-anti-Stokes coincidence probability $p_{SX,TX}^{(\text{ith})}$ etc. We now define $p_T^{(\text{ith})} = p_{TX}^{(\text{ith})} + p_{TY}^{(\text{ith})}$ as the probability of detecting an anti-Stokes photon from the $i$th source at the detector $D_T$ (either $D_T^{(1)}$ or $D_T^{(2)}$) for the non-multiplexed case, where $p_{TX}^{(\text{ith})}$ ($p_{TY}^{(\text{ith})}$) is the probability of detecting an anti-Stokes photon at the detector $D_T^{(1)}$ ($D_T^{(2)}$) set to measure the *X*-polarized (*Y*-polarized) photon for the case that only the $i$th source is operated.

Considering the background noise and imperfect polarization compensations in the detection channels, we write the Stokes and anti-Stokes detection probabilities, as well as Stokes-anti-Stokes coincidence probability for the *H-V* polarization setting as

$$p_{SH}^{(\text{ith})} = \chi_i \cos^2\vartheta(1-a_{H-V}^{(\text{ith})})\eta_{S_i} + G\eta_{S_i} + a_{H-V}^{(\text{ith})}\chi_i \sin^2\vartheta\,\eta_{S_i}, \tag{S39a}$$

$$p_{SV}^{(\text{ith})} = \chi_i \sin^2\vartheta\left(1-a_{H-V}^{(\text{ith})}\right)\eta_{S_i} + G\eta_{S_i} + a_{H-V}^{(\text{ith})}\chi_i \cos^2\vartheta\,\eta_{S_i}, \tag{S39b}$$

$$p_{TH}^{(ith)} = \chi_i \cos^2 \vartheta (1-b_{H-V}^{(ith)}) \gamma_i \eta_{T_i} + G\eta_{T_i} + b_{H-V}^{(ith)} \chi_i \sin^2 \vartheta \eta_{T_i},\tag{S39c}$$

$$p_{TV}^{(ith)} = \chi_i \sin^2 \vartheta (1-b_{H-V}^{(ith)}) \gamma_i \eta_{T_i} + G\eta_{T_i} + b_{H-V}^{(ith)} \chi_i \cos^2 \vartheta \eta_{T_i},\tag{S39d}$$

$$p_{SH,TH}^{(ith)} = \chi_i \cos^2 \vartheta \gamma_i \eta_{S_i} \eta_{T_i} (1-a_{H-V}^{(ith)} - b_{H-V}^{(ith)}) + p_{SH}^{(ith)} p_{TH}^{(ith)},\tag{S39e}$$

$$p_{SV,TV}^{(ith)} = \chi_i \sin^2 \vartheta \gamma_i \eta_{S_i} \eta_{T_i} (1-a_{H-V}^{(ith)} - b_{H-V}^{(ith)}) + p_{SV}^{(ith)} p_{TV}^{(ith)},\tag{S39f}$$

$$p_{SH,TV}^{(ith)} = (a_{H-V}^{(ith)} + b_{H-V}^{(ith)}) \chi_i \eta_{S_i} \eta_{T_i} \gamma_i \cos^2 \vartheta + p_{SH}^{(ith)} p_{TV}^{(ith)},\tag{S39g}$$

$$p_{SV,TH}^{(ith)} = (a_{H-V}^{(ith)} + b_{H-V}^{(ith)}) \chi_i \eta_{S_i} \eta_{T_i} \gamma_i \sin^2 \vartheta + p_{SV}^{(ith)} p_{TH}^{(ith)},\tag{S39h}$$

where, $G$ is the background noise which is assumed to be the same for the different channels $i = 1, 2, ..., 6$, $a_{H-V}^{(ith)} \ll 1$ ($b_{H-V}^{(ith)} \ll 1$) is the crosstalk coefficient between the $H$ and $V$ polarization components in the $i$th Stokes (anti-Stokes) channel.

For the $R$-$L$ polarization setting, the above-mentioned probabilities are written as

$$p_{SR}^{(ith)} = p_{SL}^{(ith)} = \chi_i \eta_{S_i}/2 + G\eta_{S_i},\tag{S40a}$$

$$p_{TR}^{(ith)} = p_{TL}^{(ith)} = \chi_i \gamma_i \eta_{T_i}/2 + G\eta_{T_i},\tag{S40b}$$

$$p_{SR,TR}^{(ith)} \approx \frac{1}{4}\Delta_\vartheta^2 \chi_i \gamma_i \eta_{T_i} \eta_{S_i} (1-a_{R-L}^{(ith)} - b_{R-L}^{(ith)}) + p_{SR}^{(ith)} p_{TR}^{(ith)},\tag{S40c}$$

$$p_{SL,TL}^{(ith)} \approx \frac{1}{4}\Delta_\vartheta^2 \chi_i \gamma_i \eta_{T_i} \eta_{S_i} (1-a_{R-L}^{(ith)} - b_{R-L}^{(ith)}) + p_{SL}^{(ith)} p_{TL}^{(ith)},\tag{S40d}$$

$$p_{SR,TL}^{(ith)} = \frac{1}{4}S_\vartheta^2 \chi_i \gamma_i \eta_{T_i} \eta_{S_i} + \frac{1}{4}(a_{R-L}^{(ith)} + b_{R-L}^{(ith)})\Delta_\vartheta^2 \chi_i \eta_{T_i} \eta_{S_i} \gamma_i + p_{SR}^{(ith)} p_{TL}^{(ith)},\tag{S40e}$$

$$p_{SL,TR}^{(ith)} = \frac{1}{4}S_\vartheta^2 \chi_i \gamma_i \eta_{T_i} \eta_{S_i} + \frac{1}{4}(a_{R-L}^{(ith)} + b_{R-L}^{(ith)})\Delta_\vartheta^2 \chi_i \eta_{T_i} \eta_{S_i} \gamma_i + p_{ST}^{(ith)} p_{TR}^{(ith)},\tag{S40f}$$

where, $\Delta_\vartheta = \cos\vartheta - \sin\vartheta$, $S_\vartheta = \cos\vartheta + \sin\vartheta$. $a_{R-L}^{(ith)} \ll 1$ ($b_{R-L}^{(ith)} \ll 1$) is the crosstalk coefficient between the $R$ and $L$ polarization components in the $i$th Stokes (anti-Stokes) channel. We also define the probabilities in the $D$-$A$ polarization setting by

$$p_{SD}^{(ith)} = p_{SA}^{(ith)} = \chi_i \eta_{S_i}/2 + G\eta_{S_i}, \tag{S41a}$$

$$p_{TD}^{(ith)} = p_{TA}^{(ith)} = \chi_i \gamma_i \eta_{T_i}/2 + G\eta_{T_i}, \tag{S41b}$$

$$p_{SD,TD}^{(ith)} \approx \frac{1}{4} S_\vartheta^2 \chi_i \gamma_i \eta_{T_i} \eta_{S_i} (1 - a_{D-A}^{(ith)} - b_{D-A}^{(ith)}) + p_{SD}^{(ith)} p_{TD}^{(ith)}, \tag{S41c}$$

$$p_{SA,TA}^{(ith)} \approx \frac{1}{4} S_\vartheta^2 \chi_i \gamma_i \eta_{T_i} \eta_{S_i} (1 - a_{D-A}^{(ith)} - b_{D-A}^{(ith)}) + p_{SA}^{(ith)} p_{TA}^{(ith)}, \tag{S41d}$$

$$p_{SD,TA}^{(ith)} = \frac{1}{4} \Delta_\vartheta^2 \chi_i \gamma_i \eta_{T_i} \eta_{S_i} + \frac{1}{4}(a_{D-A}^{(ith)} + b_{D-A}^{(ith)}) S_\vartheta^2 \chi_i \eta_{T_i} \eta_{S_i} \gamma_i + p_{SD}^{(ith)} p_{TA}^{(ith)}, \tag{S41e}$$

$$p_{SA,TD}^{(ith)} = \frac{1}{4} \Delta_\vartheta^2 \chi_i \gamma_i \eta_{T_i} \eta_{S_i} + \frac{1}{4}(a_{D-A}^{(ith)} + b_{D-A}^{(ith)}) S_\vartheta^2 \chi_i \eta_{T_i} \eta_{S_i} \gamma_i + p_{SA}^{(ith)} p_{TD}^{(ith)}, \tag{S41f}$$

where, $a_{D-A}^{(ith)} \ll 1$ ($b_{D-A}^{(ith)} \ll 1$) is the crosstalk coefficients between the *D* and *A* polarization Stokes (anti-Stokes) channels. Substituting the Eq. (S39), Eq. (S40) and Eq. (S41) into the Eq. (S30), we obtain the dependence of polarization visibility on the excitation probability for one SWPE source (non-multiplexed interface), e.g., the source *i*, which is

$$V_{X-Y}^{(ith)} \approx (1 - Z_{X-Y}^{(ith)}) - K_{X-Y}^{(ith)} \chi_i, \tag{S42}$$

where, *X-Y* represents the *H-V*, *R-L* or *D-A* polarization settings,

$$Z_{H-V}^{(ith)} = 2a_{H-V}^{(ith)} + 2b_{H-V}^{(ith)} + 2G(1 + 1/\gamma_i), \tag{S43a}$$

$$K_{H-V}^{(ith)} \approx (1 - 2a_{H-V}^{(ith)} - 2b_{H-V}^{(ith)}), \tag{S43b}$$

$$Z_{R-L}^{(ith)} = \left[\Delta_\vartheta^2 + 2(a_{R-L}^{(ith)} + b_{R-L}^{(ith)}) + 2G(1 + 1/\gamma_i)\right], \tag{S43c}$$

$$K_{R-L}^{(ith)} \approx \left[1 - \Delta_\vartheta^2 - 2(a_{R-L}^{(ith)} + b_{R-L}^{(ith)}) - 4G(1 + 1/\gamma_i)\right], \tag{S43d}$$

$$Z_{D-A}^{(ith)} = \left[\Delta_\vartheta^2 + 2(a_{D-A}^{(ith)} + b_{D-A}^{(ith)}) + 2G(1 + 1/\gamma_i)\right], \tag{S43e}$$

$$K_{D-A}^{(ith)} \approx \left[1 - \Delta_\vartheta^2 - 2(a_{D-A}^{(ith)} + b_{D-A}^{(ith)}) - 4G(1 + 1/\gamma_i)\right]. \tag{S43f}$$

Neglecting the noise and the cross-talks, we can write the Stokes detection rate for the source *i* as

$$R_{X-Y}^{(ith)} \approx r\chi_i \eta_{S_i}. \tag{S44}$$

And, we can write the coincidence count rate $C_{X-Y}^{(ith)}$ in the three polarization settings as

$$C_{H-V}^{(ith)} = r\left(p_{SH,TH}^{(ith)} + p_{SV,TV}^{(ith)}\right) = r\chi_i\gamma_i\eta_{S_i}\eta_{T_i}, \tag{S45a}$$

$$C_{D-A}^{(ith)} = r\left(p_{SD,TD}^{(ith)} + p_{SA,TA}^{(ith)}\right) = \frac{1}{2}rS_\vartheta^2\chi_i\gamma_i\eta_{T_i}\eta_{S_i}, \tag{S45b}$$

$$C_{R-L}^{(ith)} = r\left(p_{SR,TL}^{(ith)} + p_{SL,TR}^{(ith)}\right) = \frac{1}{2}rS_\vartheta^2\chi_i\gamma_i\eta_{T_i}\eta_{S_i}. \tag{S45c}$$

For the case of Rb atoms, $\vartheta \approx 0.81 \times \pi/4$ [3] and thus we have $S_\vartheta^2 = (\cos\vartheta + \sin\vartheta)^2 \approx 2$. So, we can express the coincidence count rate in the *X-Y* (=*H-V*, *D-A* or *R-L*) polarization setting as

$$C_{X-Y}^{(ith)} \approx r\chi_i\gamma_i\eta_{S_i}\eta_{T_i}. \tag{S46}$$

Based on the Eqs. (S42, S44 and S46), we have

$$V_{X-Y}^{(ith)} \approx \left(1 - Z_{X-Y}^{(ith)}\right) - \frac{K_{X-Y}^{(ith)}}{\Gamma_i}R_{X-Y}^{(ith)} \approx \left(1 - Z_{X-Y}^{(ith)}\right) - \frac{K_{X-Y}^{(ith)}}{\Upsilon_i}C_{X-Y}^{(ith)}, \tag{S47}$$

where, $\Gamma_i = r\eta_{S_i}$ and $\Upsilon_i = r\eta_{S_i}\gamma_i\eta_{T_i}$.

We define the average polarization visibility and the average excitation probability for the *m* sources as $\bar{V}_{X-Y} = \left(\sum_{i=1}^{m} V_{X/Y}^{(ith)}\right)/m$ and $\bar{\chi} = \left(\sum_{i=1}^{m}\chi_i\right)/m$, respectively. Also we define the average parameters as $\bar{Z}_{X-Y} = \left(\sum_{i=1}^{m} Z_{X/Y}^{(ith)}\right)/m$ and $\bar{K}_{X-Y} = \left(\sum_{i=1}^{m} K_{X-Y}^{(ith)}\right)/m$, respectively. If the deviations $\delta\chi_i = \chi_i - \bar{\chi}$, $\delta Z_{X-Y}^{(ith)} = Z_{X-Y}^{(ith)} - \bar{Z}_{X-Y}$ and $\delta K_{X-Y}^{(ith)} = K_{X-Y}^{(ith)} - \bar{K}_{X-Y}$ are very small, we can obtain

$$\bar{V}_{X-Y} \approx \left(1 - \bar{Z}_{X-Y}\right) - \bar{K}_{X-Y}\bar{\chi}. \tag{S48}$$

Furthermore, we define the average retrieval efficiency as $\bar{\gamma} = \left(\sum_{i=1}^{m}\gamma_i\right)/m$, the average detection efficiencies as $\bar{\eta}_{S(T)} = \left(\sum_{i=1}^{m}\eta_{S_i(T_i)}\right)/m$, the average Stokes detection rate as $\bar{R}_{X-Y} = \left(\sum_{i=1}^{m} r\chi_i \eta_{S_i}\right)/m$ and the average coincidence count rate as $\bar{C}_{X-Y} = \left(\sum_{i=1}^{m} C_{X-Y}^{(ith)}\right)/m \approx \left(\sum_{i=1}^{m} r\chi_i \gamma_i \eta_{S_i} \eta_{T_i}\right)/m$. If the deviations $\delta\chi_i = \chi_i - \bar{\chi}$, $\delta\gamma_i = \gamma_i - \bar{\gamma}$ and $\delta\eta_{S_i(T_i)} = \eta_{S_i(T_i)} - \bar{\eta}_{S(T)}$ are very small, we have $\bar{R}_{X-Y} = r\bar{\chi}\bar{\eta}_S$ and $\bar{C}_{X-Y} = r\bar{\chi}\bar{\gamma}\bar{\eta}_S\bar{\eta}_T$ and then rewrite the Eq. (S48) as

$$\bar{V}_{X-Y} \approx \left(1 - \bar{Z}_{X-Y}\right) - \frac{\bar{K}_{X-Y}}{\bar{\Gamma}} \cdot \bar{R}_{X-Y} \approx \left(1 - \bar{Z}_{X-Y}\right) - \frac{\bar{K}_{X-Y}}{\bar{\Upsilon}} \cdot \bar{C}_{X-Y}, \tag{S49}$$

where, $\bar{\Gamma} = r\bar{\eta}_S$ and $\bar{\Upsilon} = r \cdot \bar{\gamma} \bar{\eta}_S \bar{\eta}_T$.

**(2) Multiplexed case**

We now give the dependence of the compositive polarization visibility $V_{H-V}^{(M)}$ on the total Stokes detection rate $R_{X-Y}^{(m)}$ and the total coincidence count rate $C_{X-Y}^{(m)}$. As mentioned above, the compositive polarization visibility $V_{H-V}^{(M)}$ is defined by

$$V_{X-Y}^{(M)} = \sum_{i=1}^{m} P_{S,T}^{(M,ith)} V_{X-Y}^{(M,ith)}. \tag{S50}$$

Defining the average normalized coincidence count as $\overline{P_{S,T}^{(M)}} = \left(\sum_{i=1}^{m} P_{S,T}^{(M,ith)}\right)/m$ and average compositive polarization visibility as $\overline{V_{X-Y}^{(M)}} = \left(\sum_{i=1}^{m} V_{X-Y}^{(M,ith)}\right)/m$. According to the definition of the normalized coincidence count, we have $\sum_{i=1}^{m} P_{S,T}^{(M,ith)} = 1$. So we obtain $\overline{P_{S,T}^{(M)}} = \left(\sum_{i=1}^{m} P_{S,T}^{(M,ith)}\right)/m = 1/m$. The compositive polarization visibility $V_{X-Y}^{(M)}$ can be expressed as

$$V_{X-Y}^{(M)} = \overline{V_{X-Y}^{(M)}} + \sum_{i=1}^{m} \delta P_{S,T}^{(M,ith)} \cdot \delta V_{X-Y}^{(M,ith)}, \tag{S51}$$

where, $\delta P_{S,T}^{(M,ith)} = P_{S,T}^{(M,ith)} - \overline{P_{S,T}^{(M)}} = P_{S,T}^{(M,ith)} - 1/m$ and $\delta V_{X-Y}^{(M,ith)} = V_{X-Y}^{(M,ith)} - \overline{V_{X-Y}^{(M)}}$. Assuming that the deviations $\delta P_{S,T}^{(M,ith)}$ and $\delta V_{X-Y}^{(M,ith)}$ are very small and the quantity $\sum_{i=1}^{m} \delta P_{S,T}^{(M,ith)} \cdot \delta V_{X-Y}^{(M,ith)}$ can

be neglected, we obtain

$$V_{X-Y}^{(M)} \approx \overline{V_{X-Y}^{(M)}} . \tag{S52}$$

The average compositive polarization visibility $\overline{V_{X-Y}^{(M)}}$ can be rewritten as

$$\overline{V_{X-Y}^{(M)}} = \left(\sum_{i=1}^{m} V_{X-Y}^{(M,ith)}\right)/m = \frac{1}{m}\left(\sum_{i=1}^{m} \frac{C_{X-Y}^{(M,ith)} - N_{X-Y}^{(M,ith)}}{C_{X-Y}^{(M,ith)} + N_{X-Y}^{(M,ith)}}\right)$$
$$= \frac{1}{m}\sum_{i=1}^{m} \frac{\left| p_{SX,TY}^{(M,ith)} + p_{SY,TX}^{(M,ith)} - p_{SX,TY}^{(M,ith)} - p_{SY,TX}^{(M,ith)} \right|}{p_{SX,TY}^{(M,ith)} + p_{SY,TX}^{(M,ith)} + p_{SX,TY}^{(M,ith)} + p_{SY,TX}^{(M,ith)}} . \tag{S53}$$

In Eq. (S27), we have obtained the relations: $p_{SX,TX}^{(M,ith)} \approx \eta_{RC_i} p_{SX,TX}^{(ith)}$, $p_{SY,TY}^{(M,ith)} \approx \eta_{RC_i} p_{SY,TY}^{(ith)}$, $p_{SX,TY}^{(M,ith)} \approx \eta_{RC_i} p_{SX,TY}^{(ith)}$, $p_{SY,TX}^{(M,ith)} \approx \eta_{RC_i} p_{SY,TX}^{(ith)}$ for the case of $m\overline{p}_S \ll 1$. Based these relations, we can rewrite the Eq. (S53) as

$$\overline{V_{X-Y}^{(M)}} \approx \frac{1}{m}\sum_{i=1}^{m} \frac{\left| p_{SX,TY}^{(ith)} + p_{SY,TX}^{(ith)} - p_{SX,TY}^{(ith)} - p_{SY,TX}^{(ith)} \right|}{p_{SX,TY}^{(ith)} + p_{SY,TX}^{(ith)} + p_{SX,TY}^{(ith)} + p_{SY,TX}^{(ith)}} = \overline{V}_{X-Y} . \tag{S54}$$

Based on the above equation, we can rewrite the expression $V_{X-Y}^{(M)}$ as

$$V_{X-Y}^{(M)} \approx \overline{V_{X-Y}^{(M)}} \approx \overline{V}_{X-Y} = \left(\sum_{i=1}^{m} V_{X-Y}^{(ith)}\right)/m = \left(1 - \overline{Z}_{X-Y}\right) - \overline{K}_{X-Y}\overline{\chi} . \tag{S55}$$

The total Stokes detection rate $R_{X-Y}^{(m)}$ expressed in Eq. (S23) can be expressed as

$$R_{X-Y}^{(m)} \approx r\sum_{i=1}^{m} \chi_i \eta_{S_i} \approx mr\overline{\eta}_S \overline{\chi} = m\overline{\Gamma}\overline{\chi} . \tag{S56}$$

Substituting the Eq. (S56) into the Eq. (S55), we obtain the dependence of the compositive polarization visibility $V_{H-V}^{(M)}$ on the total Stokes detection rate $R_{X-Y}^{(m)}$, which is

$$V_{X-Y}^{(M)} \approx \left(1 - \overline{Z}_{X-Y}\right) - \frac{1}{m}\frac{\overline{K}_{X-Y} R_{X-Y}^{(m)}}{\overline{\Gamma}} . \tag{S57}$$

Based on the Eq. (S28), $C_{X-Y}^{(m)}$ can be expressed as $C_{X-Y}^{(m)} = \sum_{i=1}^{m} C_{X-Y}^{(M,ith)} \approx \sum_{i=1}^{m} \eta_{RC_i} C_{X-Y}^{(ith)}$.

From the Eq. (S46), we have

$$C_{X-Y}^{(m)} \approx r\sum_{i=1}^{m}\chi_i\eta_{RC_i}\gamma_i\eta_{S_i}\eta_{T_i} \approx r\bar{\eta}_{RC}\sum_{i=1}^{m}\chi_i\gamma_i\eta_{S_i}\eta_{T_i} = m\bar{\eta}_{RC}\bar{\chi}\cdot\bar{\Upsilon}. \qquad (S58)$$

Substituting Eq. (S58) into the Eq. (S55), we obtain the dependence of $V_{H-V}^{(M)}$ on the total coincidence count rate $C_{X-Y}^{(m)}$, which is

$$V_{X-Y}^{(M)} \approx \left(1-\bar{Z}_{X-Y}\right) - \frac{1}{m\bar{\eta}_{RC}}\frac{\bar{K}_{X-Y}C_{X-Y}^{(m)}}{\bar{\Upsilon}}. \qquad (S59)$$

The Eq. (S57) [Eq. (S59)] shows that the multiplexed interface enables an $m$-fold [$m\bar{\eta}_{RC}$-fold] increase in the Stokes detection rate [coincidence count rate] compared to the non-multiplexed one [Eq. (S47)] for a fixed polarization visibility.

**VIII. The success probabilities for entanglement generation in a single link**:

We first discuss the generation rates of an entangled state in a single link using the non-multiplexed interfaces and then discuss that using the multiplexed interfaces.

(1) **For the non-multiplexed case,** the two spin-wave-photon entanglement (SWPE) sources, for example, the source $|\Phi\rangle_A^{(ith)}$ in the ensemble A and the source $|\Phi\rangle_B^{(ith)}$ in the ensemble B, are considered. i.e., we pay attention to the BSM of the Stokes photons $S_{A_i}$ and $S_{B_i}$ at the $i$th station.

The atom-photon entangled states generated from the sources can be written as [3, 4]:

$$|\Phi\rangle_A^{(ith)} = \left(\cos\vartheta|+\rangle_A^{(ith)}|R\rangle_A^{(ith)} + \sin\vartheta|-\rangle_A^{(ith)}|L\rangle_A^{(ith)}\right), \qquad (S60a)$$

$$|\Phi\rangle_B^{(ith)} = \left(\cos\vartheta|+\rangle_B^{(ith)}|R\rangle_B^{(ith)} + \sin\vartheta|-\rangle_B^{(ith)}|L\rangle_B^{(ith)}\right). \qquad (S60b)$$

The success probability to detect a Stokes photon $S_{A_i}$ ($S_{B_i}$) from A (B)

ensemble is defined as $p_{S_A}^{(ith)}$ ($p_{S_B}^{(ith)}$), which can be expressed as $p_{S_A}^{(ith)} = \chi_{A_i} \eta_{S_{A_i}}$ ($p_{S_B}^{(ith)} = \chi_{B_i} \eta_{S_{B_i}}$), where $\chi_{A_i}$ ($\chi_{B_i}$) is the excitation probability of the $i$th source in the ensemble A (B) per write pulse, $\eta_{S_{A_i}}$ ($\eta_{S_{B_i}}$) is the total detection efficiency for detecting a Stokes photon at $i$th channel of the A (B) ensemble.

For convenience, we place a $\lambda/4$ plate in the optical paths of the Stokes photons $S_{A_i}$ ($S_{B_i}$) from the source A (B) in order to transform its $\sigma^+(\sigma^-)$ polarization into $H(V)$ [$V(H)$] polarization. Thus the SWPE states are rewritten as:

$$|\Phi\rangle_A^{(ith)} = \left(\cos\vartheta |+\rangle_A^{(ith)} |H\rangle_A^{(ith)} + \sin\vartheta |-\rangle_A^{(ith)} |V\rangle_A^{(ith)}\right), \quad (S61a)$$

$$|\Phi\rangle_B^{(ith)} = \left(\cos\vartheta |+\rangle_B^{(ith)} |H\rangle_B^{(ith)} + \sin\vartheta |-\rangle_B^{(ith)} |V\rangle_B^{(ith)}\right), \quad (S61b)$$

respectively. For obtaining a maximal entangled state between A and B ensembles (cf. below), we transform the state $|\Phi\rangle_B^{(ith)} = \left(\cos\vartheta |+\rangle_B^{(ith)} |H\rangle_B^{(ith)} + \sin\vartheta |-\rangle_B^{(ith)} |V\rangle_B^{(ith)}\right)$ into the state $|\Phi\rangle_B^{(ith)} = \left(\cos\vartheta |+\rangle_B^{(ith)} |V\rangle_B^{(ith)} + \sin\vartheta |-\rangle_B^{(ith)} |H\rangle_B^{(ith)}\right)$ by placing a $\lambda/2$ plate in the path of the Stokes photons $S_{A_i}$ to transform $|H\rangle_B^{(ith)} / |V\rangle_B^{(ith)}$ into $|V\rangle_B^{(ith)} / |H\rangle_B^{(ith)}$. The Stokes photons $S_{A_i}$ and $S_{B_i}$ are collected by two single-mode fibers, respectively. For reducing transmission losses in the optical fibers, we may convert the Stokes photons at 795nm (Rb atomic transition wavelength) to the telecom C-band and then sent them to the station. The two photons are combined on a polarization-beam-splitter (PBS) at the $i$th station for BSM. If the two Stokes photons are in the same polarization $H$ or $V$, they will exit from two different output ports of the PBS. Thus, a four-particle entangled state will be formed

[5], which is written as:

$$|\Phi\rangle_{AB}^{(ith)} = \frac{\chi \sin 2\vartheta}{\sqrt{2}}\left(|+\rangle_A^{(ith)}|-\rangle_B^{(ith)}|H\rangle_A^{(ith)}|H\rangle_B^{(ith)} + |-\rangle_A^{(ith)}|+\rangle_B^{(ith)}|V\rangle_A^{(ith)}|V\rangle_B^{(ith)}\right). \quad (S62)$$

Since the probability that the two Stokes photons are in the same polarization is only 0.5, the BSM efficiency is 50%. For the Rb atom system, it is reasonable to assume $\sin 2\vartheta \approx 1$ [3]. Neglecting the vacuum state, the four-particle entangled state $|\Phi\rangle_{AB}^{(ith)}$ can be expressed as:

$$|\Phi\rangle_{AB}^{(ith)} = \frac{1}{2}\left(|\Phi^+\rangle_{AB}^{(ith)}|\varphi^+\rangle_{AB}^{(ith)} + |\Phi^-\rangle_{AB}^{(ith)}|\phi^-\rangle_{AB}^{(ith)}\right), \quad (S63)$$

where, $|\Phi^\pm\rangle_{AB}^{(ith)} = \frac{1}{\sqrt{2}}\left(|+\rangle_A^{(ith)}|-\rangle_B^{(ith)} \pm |-\rangle_A^{(ith)}|+\rangle_B^{(ith)}\right)$, $|\varphi^+\rangle_{AB}^{(ith)} = \left(|D\rangle_A^{(ith)}|D\rangle_B^{(ith)} + |A\rangle_A^{(ith)}|A\rangle_B^{(ith)}\right)/\sqrt{2}$ and $|\phi^-\rangle_{AB}^{(ith)} = \left(|D\rangle_A^{(ith)}|A\rangle_B^{(ith)} + |A\rangle_A^{(ith)}|D\rangle_B^{(ith)}\right)/\sqrt{2}$ are the Bell states, $|D\rangle = (|H\rangle + |V\rangle)/\sqrt{2}$ and $|A\rangle = (|H\rangle - |V\rangle)/\sqrt{2}$ correspond to $+45°$ and $-45°$ linear polarizations, respectively. The Bell state $|\varphi^+\rangle_{AB}^{(ith)}$ ($|\phi^-\rangle_{AB}^{(ith)}$) can be identified by a suitable joint polarization measurement on the photons $s_{A_i}$ and $s_{B_i}$. According to the measurement outcomes of the BSM [6], one can identify that the memories in the A and B ensembles are projected onto the maximal entangled state:

$$|\Phi^+\rangle_{AB}^{(ith)} = \frac{1}{\sqrt{2}}\left(|+\rangle_A^{(ith)}|-\rangle_B^{(ith)} + |-\rangle_A^{(ith)}|+\rangle_B^{(ith)}\right), \quad (S64a)$$

$$\text{or } |\Phi^-\rangle_{AB}^{(ith)} = \frac{1}{\sqrt{2}}\left(|+\rangle_A^{(ith)}|-\rangle_B^{(ith)} - |-\rangle_A^{(ith)}|+\rangle_B^{(ith)}\right). \quad (S64b)$$

If it is the entangled state $|\Phi^-\rangle_{AB}^{(ith)}$, one can apply a linear operation [7] on the relative phase between the memory qubit in the ensemble A (or B) to transform it into the entangled state $|\Phi^+\rangle_{AB}^{(ith)}$. So, the probability to successfully generate the entangled state $|\Phi^+\rangle_{AB}^{(ith)}$ is on the order of [8, 9]

$$p_{S_{L0}}^{(ith)} = \frac{1}{2} p_{S_{AB}}^{(ith)} e^{-L_0/L_{att}} = \frac{1}{2} p_{S_A}^{(ith)} p_{S_B}^{(ith)} e^{-L_0/L_{att}}, \quad (S65)$$

where, $L_{att} = 22\text{km}$ is the fiber attenuation length, 1/2-factor comes from the BSM efficiency of 50% mentioned above.

Defining the average success probability for entanglement generation over the $m$-channels as $\bar{p}_{S_{L0}} = \left(\sum_{i=1}^{m} p_{S_{L0}}^{(i\text{th})}\right)/m$ and assuming that the relative deviation $\delta p_{S_{L0}}^{(i\text{th})} / \bar{p}_{S_{L0}} = \left(p_{S_{L0}}^{(i\text{th})} - \bar{p}_{S_{L0}}\right)/\bar{p}_{S_{L0}} \ll 1$, we have

$$p_{S_{L0}}^{(1\text{st})} \approx p_{S_{L0}}^{(2\text{nd})} \approx ... \approx p_{S_{L0}}^{(m\text{th})} \approx \bar{p}_{S_{L0}} = p_{S_{L0}}^{(1)}, \tag{S66}$$

meaning that the success probability for entanglement generation between the ensembles A and B for the non-multiplexed case, i.e., each ensemble using one source, is $p_{S_{L0}}^{(1)}$.

As mentioned above, we have to convert the frequency of the Stokes photons from 795 nm to the telecom wavelength for reducing the transmission loss of the Stokes photon in the optical fibers. To implement such quantum frequency conversion, we may use a quantum frequency converter device (QFCD) demonstrated in Ref. [1]. The total efficiency of the QFCD, including the internal conversion efficiency of the nonlinear periodically-poled-lithium-niobate (PPLN) waveguide and the overall efficiency of optical elements, such as Bragg grating filter, optical fiber coupler, etc., is non-unity, which will affect the entanglement generation rate in each elementary link. Here, we assumes that the total efficiencies of the QFCDs used at the $i$th Stokes channels of the ensembles A and B are all $\eta_{DC}$. After the imperfect total efficiency $\eta_{DC}$ is taken into account, the success probability $p_{S_{L0}}^{(1)}$ will become:

$$p_{S_{L0}}^{(1)}{}' = p_{S_{L0}}^{(i\text{th})}{}' = \frac{1}{2} p_{S_A}^{(i\text{th})} p_{S_B}^{(i\text{th})} \eta_{DC}^2 e^{-L_0/L_{att}} = \eta_{DC}^2 p_{S_{L0}}^{(1)}, \tag{S67}$$

which shows that the success probability $p_{S_{L0}}^{(1)}$ will decrease by a factor of $\eta_{DC}^2$. Recently, the state-of-the-art conversion efficiency of the nonlinear PPLN waveguide is up to ~77.1%, while the overall efficiency of the optical elements is ~17.64% [1]. In this case, the total efficiency of the QFCD is only $\eta_{DC} \approx 77.1\% * 17.64\% \approx 13.6\%$. If such quantum frequency conversion technique is used in the presented scheme, the probability $p_{S_{L0}}^{(1)}$ will decrease by a factor of $\eta_{DC}^2 \approx 1.85\%$. Such factor is quite small, but can be significantly enhanced by technical improvements such as reducing the optical losses of the QFCD [1].

(2) **For the multiplexed case,** a spatial array of *m* SWPE sources is simultaneously excited in each ensemble (A or B), these sources generate the entangled states $|\Phi\rangle_{A(B)}^{(1st)}$, $|\Phi\rangle_{A(B)}^{(2nd)}$,…$|\Phi\rangle_{A(B)}^{(ith)}$…$|\Phi\rangle_{A(B)}^{(mth)}$ in a probabilistic manner. The SWPE source $|\Phi\rangle_A^{(ith)}$ ($|\Phi\rangle_B^{(ith)}$) can create an entangled pair of a Stokes photon $S_{A_i}$ ($S_{B_i}$) and one spin-wave (collective) excitation $M_{A_i}$ ($M_{B_i}$). The Stokes photons $S_{A_i}$ and $S_{B_i}$ are sent to a polarization-beam-splitter at the *i*th middle station via two optical fibers, respectively. For reducing transmission losses in the optical fibers, the wavelength of the Stokes photons has to be converted from 795nm to the telecom C-band before transmitting in the optical fibers. After the polarization-beam-splitter at the *i*th middle station, the two Stokes photons are performed a joint BSM. Any one of successful BSMs at the *1*st, 2nd, …*i*th, ... *m*th stations will project the A and B ensembles into the joint atom-atom state:

$$\rho_{AB}^{(M)} = p_{S_{L0}}^{(m)} \rho_{ent}^{(M)} + (1 - p_{S_{L0}}^{(m)})|0\rangle\langle 0|, \tag{S68}$$

where, the high-order excitations have been neglected, $|0\rangle\langle 0|$ denote the vacuum state, $\rho_{ent}^{(M)}$ is a mixed entangled state between the ensembles A and B, which can be written as

$$\rho_{ent}^{(M)} = \sum_{i=1}^{m} P_{S_{L0}}^{(M,ith)} \rho_{ent}^{(M,ith)} = \sum_{i=1}^{m} P_{S_{L0}}^{(M,ith)} \rho_{ent}^{(ith)}, \tag{S69}$$

$\rho_{ent}^{(M,ith)}$ represents the $i$th entangled state between the ensembles A and B generated in this multiplexed case, which is equal to that for the non-multiplexed case, i.e., $\rho_{ent}^{(M,ith)} = \rho_{ent}^{(ith)}$, with $\rho_{ent}^{(ith)} = |\Phi^+\rangle_{AB}^{(ith)}\,{}_{AB}^{(ith)}\langle\Phi^+|$, $|\Phi^+\rangle_{AB}^{(ith)} = \frac{1}{\sqrt{2}}(|+\rangle_A^{(ith)}|-\rangle_B^{(ith)} + |-\rangle_A^{(ith)}|+\rangle_B^{(ith)})$ defined in the Eq. (64a), $p_{S_{L0}}^{(m)} = \sum_{i=1}^{m} p_{S_{L0}}^{(M,ith)}$ is the total success probability for entanglement generation for the single link using the multiplexed interfaces (MIs) per write-pulse, $P_{S_{L0}}^{(M,ith)} = p_{S_{L0}}^{(M,ith)} / p_{S_{L0}}^{(m)}$ is normalized probability, $p_{S_{L0}}^{(M,ith)}$ is the probability of a successful BSM at the $i$th station for the multiplexed case,

The probability $p_{S_{L0}}^{(M,ith)}$ can be expressed as

$$p_{S_{L0}}^{(M,ith)} = \left(\prod_{l=0}^{i-1} k_l\right) p_{S_{L0}}^{(ith)}, \tag{S70}$$

where, $k_0 = 1$, $k_1 = 1 - p_{S_{L0}}^{(1st)}$, $k_2 = \left(1 - p_{S_{L0}}^{(2nd)}\right)$, ... $k_l = \left(1 - p_{S_{L0}}^{(lst)}\right)$. Assuming that $p_{S_{L0}}^{(1st)} \approx p_{S_{L0}}^{(2nd)} \approx ... \approx p_{S_{L0}}^{(mth)} \approx p_{S_{L0}}^{(1)}$, we have $\left(\prod_{l=0}^{i-1} k_l\right) \approx \left(1 - p_{S_{L0}}^{(1)}\right)^{(i-1)}$ and then give the total success probability $p_{S_{L0}}^{(m)}$ as

$$p_{S_{L0}}^{(m)} \approx 1 - \left(1 - p_{S_{L0}}^{(1)}\right)^m. \tag{S71}$$

For $m\bar{p}_{S_{L0}} \ll 1$, we rewrite Eq. (S71) as

$$p_{S_{L0}}^{(m)} \approx m p_{S_{L0}}^{(1)}, \tag{S72}$$

showing that the multiplexed scheme enables an $m$-fold increase in the probability for generating entanglement in an elementary link compared to the non-multiplexed scheme.

For reducing the transmission loss of the Stokes photon in the optical fibers, the Stokes photons from each of the $m$ sources are required to experience the quantum frequency conversion by using the mentioned-above QFCDs. The total efficiency of the QFCD is non-unity and should be taken into account in the calculation for the success probability $p_{S_{L0}}^{(m)}$. Here, we assume that the total efficiency of the QFCD used in each of the Stokes channels is $\eta_{DC}$. After considering the imperfect efficiency $\eta_{DC}(<1)$, the total success probability $p_{S_{L0}}^{(m)}$ will become:

$$p_{S_{L0}}^{(m)}{}' = \eta_{DC}^2 p_{S_{L0}}^{(m)}, \tag{S73}$$

which shows that $p_{S_{L0}}^{(m)}$ will decrease by a factor of $\eta_{DC}^2$.

The total success probability $p_{S_{L0}}^{(m)}{}'$ included the total efficiency $\eta_{DC}$ can be rewritten as

$$p_{S_{L0}}^{(m)}{}' = m\eta_{DC}^2 p_{S_{L0}}^{(1)} = m p_{S_{L0}}^{(1)}{}', \tag{S74}$$

where $p_{S_{L0}}^{(1)}{}' = \eta_{DC}^2 p_{S_{L0}}^{(1)}$ denotes the entanglement generation probability which absorbs the total efficiency $\eta_{DC}$ and is for the non-multiplexed case. The above equation shows that the multiplexed scheme still promises an $m$-fold increase in the probability for generating entanglement in an elementary link after considering the imperfect total efficiency $\eta_{DC}$.

## IX. The time required for successfully generating entanglement over a long distance $L$

After the entanglement is generated in each elementary link, one can extend entanglement via successive entanglement-swapping operations between two adjacent links [10]. The entanglement swapping operations rely on the joint BSM between the recalled photons from two adjacent memories. For the non-multiplexed and multiplexed schemes, the average total time needed for the entanglement distribution over the distance $L$ with $n$ nest-level are $T_{tot}^{(one)}$ and $T_{tot}^{(m)}$, which are given by [9]

$$T_{tot}^{(one)} \approx \frac{L}{c} \frac{1}{p_{S_{L0}}^{(1)} p_1 p_2 \cdots p_n} \left(\frac{3}{2}\right)^n, \text{ and} \tag{S75a}$$

$$T_{tot}^{(m)} \approx \frac{L}{c} \frac{1}{p_{S_{L0}}^{(m)} \bar{\eta}_{RC}^{2n} p_1 p_2 \cdots p_n} \left(\frac{3}{2}\right)^n, \tag{S75b}$$

respectively, where $p_1 = p_2 = \cdots = p_n \approx \frac{1}{2} \gamma_A \gamma_B \eta_{S_A} \eta_{S_B}$ are the success probabilities for the entanglement swapping step for the first, second, …, $n$th swapping operations, $\gamma_A$ ($\gamma_B$) is the retrieval efficiency from the A (B) ensemble, $\bar{\eta}_{RC} = \bar{\eta}_{OSN} * \bar{\eta}_{CF}$ is the transmission of the router circuitry, including the average OSN transmission $\bar{\eta}_{OSN}$ and CSMF coupling efficiency $\bar{\eta}_{CF}$, 1/2 factor is the BSM efficiency of 50%.

For $mp_{S_{L0}}^{(1)} \ll 1$, the total times for the non-multiplexed case and multiplexed case can be approximately expressed as:

$$T_{tot}^{(one)} \approx \frac{L}{c} \frac{1}{p_{S_{L0}}^{(1)} p_1 p_2 \cdots p_n} \left(\frac{3}{2}\right)^n, \text{ and} \tag{S76a}$$

$$T_{tot}^{(m)} \approx \frac{1}{m\bar{\eta}_{RC}^{2n}} \frac{L}{c} \frac{1}{p_{S_{L0}}^{(1)} p_1 p_2 \cdots p_n} \left(\frac{3}{2}\right)^n, \tag{S76b}$$

respectively, which show that the multiplexed QR protocol can reduce the total time by a factor of $m\bar{\eta}_{RC}^{2n}$.

For the scheme discussed in the main text (as shown in Fig. 1 in the main text), 1-level swapping step is required to distribute entanglement over the distance $L$, the multiplexed QR will reduce the total time by a factor of $m\bar{\eta}_{RC}^2$ over the non-multiplexed one.

If the imperfect total efficiency $\eta_{DC}$ is taken into account, the total times $T_{tot}^{(one)}$ and $T_{tot}^{(m)}$ will become:

$$T_{tot}^{(one)}{}' = \frac{T_{tot}^{(one)}}{\eta_{DC}^2} \approx \frac{L}{c} \frac{1}{\eta_{DC}^2 p_{S_{L0}}^{(1)} p_1 p_2 \cdots p_n} \left(\frac{3}{2}\right)^n, \text{ and} \quad (S77a)$$

$$T_{tot}^{(m)}{}' = \frac{T_{tot}^{(m)}}{\eta_{DC}^2} \approx \frac{1}{m\bar{\eta}_{RC}^{2n}} \frac{L}{c} \frac{1}{\eta_{DC}^2 p_{S_{L0}}^{(1)} p_1 p_2 \cdots p_n} \left(\frac{3}{2}\right)^n, \quad (S77b)$$

which show that both of them increase by a factor of $1/\eta_{DC}^2$.

## X. The dependence of the quality of the entangled state generated in an elementary link on that generated in the multiplexed interfaces

According to the definition of the entangled state $\rho_{ent}^{(M)}$ defined in Eq. (S69), we may give the visibility of the state $\rho_{ent}^{(M)}$ as

$$V_{AB}^{(M)} = \sum_{i=1}^{m} P_{S_{L0}}^{(M,ith)} V_{AB}^{(ith)}, \quad (S78)$$

where, $P_{S_{L0}}^{(M,ith)} = p_{S_{L0}}^{(M,ith)} / p_{S_{L0}}^{(m)}$ is normalized probability, $V_{AB}^{(ith)}$ is the visibility of the $i$th entangled state $\rho_{ent}^{(ith)}$ defined by the Eq. (64a), which is equal to the $i$th entangled state $\rho_{ent}^{(M,ith)}$ for the multiplexed case. The normalized probability can be rewritten as

$$P_{S_{L0}}^{(M,ith)} = \frac{p_{S_{L0}}^{(M,ith)}}{\sum_{i=1}^{m} p_{S_{L0}}^{(M,ith)}} = \frac{\frac{1}{2} e^{-L_0/L_{att}} p_{S_{AB}}^{(M,ith)}}{\frac{1}{2} e^{-L_0/L_{att}} \sum_{i=1}^{m} p_{S_{AB}}^{(M,ith)}} = \frac{p_{S_{AB}}^{(M,ith)}}{\sum_{i=1}^{m} p_{S_{AB}}^{(M,ith)}}, \quad (S79)$$

where $p_{S_{AB}}^{(M,ith)} = \left(1 - \bar{p}_{S_{L0}}\right)^{(i-1)} p_{S_A}^{(ith)} p_{S_B}^{(ith)}$. For $m\bar{p}_{S_{L0}} \ll 1$, we have $\left(1 - \bar{p}_{S_{L0}}\right)^{(i-1)} \approx 1$, and then rewrite the $V_{AB}^{(M)}$ as

$$V_{AB}^{(M)} \approx \frac{\sum_{i=1}^{m} p_{S_A}^{(ith)} p_{S_B}^{(ith)} V_{AB}^{(ith)}}{\sum_{i=1}^{m} p_{S_A}^{(ith)} p_{S_B}^{(ith)}}. \tag{S80}$$

The visibility $V_{AB}^{(ith)}$ of the entangled state $\rho_{ent}^{(ith)}$ depends on the visibilities of the spin-wave-photon entangled states $|\Phi\rangle_A^{(ith)}$ and $|\Phi\rangle_B^{(ith)}$, which can be expressed as [10, 11]

$$V_{AB}^{(ith)} = \zeta_i V_A^{(ith)} V_B^{(ith)}, \tag{S81}$$

where, $V_A^{(ith)}$ ($V_B^{(ith)}$) is the visibility of the SWPE state $|\Phi\rangle_A^{(ith)}$ ($|\Phi\rangle_B^{(ith)}$) in the ensemble A (B), $\zeta_i$ is the parameter related to the reliability of the entanglement swapping operation [11] and assumed to be $\zeta_i = 1$ in the present work. So, we can rewrite $V_{AB}^{(M)}$ as

$$V_{AB}^{(M)} = \frac{1}{\sum_{i=1}^{m} p_{S_A}^{(ith)} p_{S_B}^{(ith)}} \sum_{i=1}^{m} p_{S_A}^{(ith)} V_A^{(ith)} p_{S_B}^{(ith)} V_B^{(ith)}. \tag{S82}$$

The visibility $V_{A(B)}^{(ith)}$ can be obtained by measuring the visibility of the entangled pair of the Stokes $S_{A(B)_i}$ and anti-Stokes photons $T_{A(B)_i}$.

We now define the compositive visibilities of the multiplexed interfaces A and B as

$$V_A^{(M)} = \sum_{i=1}^{m} P_{S_A,T_A}^{(M,ith)} V_A^{(M,ith)}, \tag{S83a}$$

$$V_B^{(M)} = \sum_{i=1}^{m} P_{S_B,T_B}^{(M,ith)} V_B^{(M,ith)}, \tag{S83b}$$

respectively, $P_{S_A,T_A}^{(M,ith)} = \frac{p_{S_A,T_A}^{(M,ith)}}{\sum_{i=1}^{m} p_{S_A,T_A}^{(M,ith)}}$ ($P_{S_B,T_B}^{(M,ith)} = \frac{p_{S_A,T_A}^{(M,ith)}}{\sum_{i=1}^{m} p_{S_A,T_A}^{(M,ith)}}$) is the normalized coincidence probability, with $\sum_{i=1}^{m} P_{S_A,T_A}^{(M,ith)} = \sum_{i=1}^{m} P_{S_B,T_B}^{(M,ith)} = 1$, $p_{S_A,T_A}^{(M,ith)} = (1-\bar{p}_{S_A})^{i-1} \eta_{RC_{A_i}} p_{S_A,T_A}^{(ith)}$ [$p_{S_B,T_B}^{(M,ith)} = (1-\bar{p}_{S_B})^{i-1} \eta_{RC_{B_i}} p_{S_B,T_B}^{(ith)}$] is the coincidence probability of detecting a pair of photons, one is at Stokes channel and another is at anti-Stokes channel of the source $i$ in the A (B) ensemble for the multiplexed case, $p_{S_A,T_A}^{(ith)}$ ($p_{S_B,T_B}^{(ith)}$) is the corresponding coincidence probability measured for the non-multiplexed case, $\bar{p}_{S_A} = \frac{1}{m}\sum_{i=1}^{m} p_{S_A}^{(ith)} = \frac{1}{m}\sum_{i=1}^{m} \chi_{A_i}\eta_{S_{A_i}}$ ($\bar{p}_{S_B} = \frac{1}{m}\sum_{i=1}^{m} p_{S_B}^{(ith)} = \frac{1}{m}\sum_{i=1}^{m} \chi_{B_i}\eta_{S_{B_i}}$) is the average Stokes detection probability, $p_{S_A}^{(ith)}$ ($p_{S_B}^{(ith)}$) is the probability of detecting a photon at the Stokes channel of the $i$th source in the A (B) ensemble, $\chi_{A_i}$ ($\chi_{B_i}$) is the excitation probability of the $i$th source in the A (B) ensemble, $\eta_{S_{A_i}}$ ($\eta_{S_{B_i}}$) denotes the total efficiency for detecting a Stokes photon at the Stokes channel of the $i$th source in the A (B) ensemble, $V_A^{(M,ith)}$ ($V_B^{(M,ith)}$) is the visibility of the entangled Stokes and anti-Stokes photons from the source $i$ in the ensemble A (B), the superscript $M$ denotes that the measurement of the $V_A^{(M,ith)}$ ($V_B^{(M,ith)}$) is for the multiplexed case.

Considering the cases of $m\bar{p}_{S_A} \ll 1$ and $m\bar{p}_{S_B} \ll 1$, we have $(1-\bar{p}_{S_A})^{i-1} \approx 1$ and $(1-\bar{p}_{S_B})^{i-1} \approx 1$. According to the Eqs. (S36 and S11), we may obtain

$$p_{S_A,T_A}^{(M,ith)} \approx \eta_{RC_{A_i}} p_{S_A,T_A}^{(ith)} \approx \eta_{RC_{A_i}} \chi_{A_i}\gamma_{A_i}\eta_{S_{A_i}}\eta_{T_{A_i}} \approx \eta_{RC_{A_i}} p_{S_A}^{(ith)}\gamma_{A_i}\eta_{T_{A_i}}, \tag{S84a}$$

$$p_{S_B,T_B}^{(M,ith)} \approx \eta_{RC_{B_i}} p_{S_B,T_B}^{(ith)} \approx \eta_{RC_{B_i}} \chi_{B_i}\gamma_{B_i}\eta_{S_{B_i}}\eta_{T_{B_i}} \approx \eta_{RC_{B_i}} p_{S_B}^{(ith)}\gamma_{B_i}\eta_{T_{B_i}}, \tag{S84b}$$

where $\gamma_{A(B)_i}$ denotes the retrieval efficiency of the anti-Stokes photon from the source $i$ of the A (B) ensemble, $\eta_{T_{A_i}}$ ($\eta_{T_{B_i}}$) denotes the total efficiency for detecting an anti-Stokes photon from the $i$th source of the A (B) ensemble. Assuming that $\eta_{RC_{A_1}} = \eta_{RC_{A_2}} = ... = \eta_{RC_{A_i}} = \eta_{RC_A}$ and $\eta_{RC_{B_1}} = \eta_{RC_{B_2}} = ... = \eta_{RC_{B_i}} = \eta_{RC_B}$, we can rewrite Eq. (S83) as:

$$V_A^{(M)} = \frac{\eta_{RC_A} \sum_{i=1}^{m} p_{S_A}^{(ith)} \gamma_{A_i} \eta_{T_{A_i}} V_A^{(M, ith)}}{\eta_{RC_A} \sum_{i=1}^{m} p_{S_A}^{(ith)} \gamma_{A_i} \eta_{T_{A_i}}} = \frac{\sum_{i=1}^{m} p_{S_A}^{(ith)} \gamma_{A_i} \eta_{T_{A_i}} V_A^{(M, ith)}}{\sum_{i=1}^{m} p_{S_A}^{(ith)} \gamma_{A_i} \eta_{T_{A_i}}}, \tag{S85a}$$

$$V_B^{(M)} = \frac{\eta_{RC_B} \sum_{i=1}^{m} p_{S_B}^{(ith)} \gamma_{B_i} \eta_{T_{B_i}} V_B^{(M, ith)}}{\eta_{RC_B} \sum_{i=1}^{m} p_{S_B}^{(ith)} \gamma_{B_i} \eta_{T_{B_i}}} = \frac{\sum_{i=1}^{m} p_{S_B}^{(ith)} \gamma_{B_i} \eta_{T_{B_i}} V_B^{(M, ith)}}{\sum_{i=1}^{m} p_{S_B}^{(ith)} \gamma_{B_i} \eta_{T_{B_i}}}. \tag{S85b}$$

Assuming that the router circuitries used in the A and B ensembles don't introduce extra noise, we can obtain that the visibility of the entangled two photons from the source $i$ in the ensemble A (B) is equal to the visibility of the SWPE state $|\Phi\rangle_A^{(ith)}$ ($|\Phi\rangle_B^{(ith)}$), i.e., $V_A^{(M, ith)} = V_A^{(ith)}$ and $V_B^{(M, ith)} = V_B^{(ith)}$. Thus, Eq. (S85) can be rewritten as:

$$V_A^{(M)} = \frac{\sum_{i=1}^{m} p_{S_A}^{(ith)} \gamma_{A_i} \eta_{T_{A_i}} V_A^{(ith)}}{\sum_{i=1}^{m} p_{S_A}^{(ith)} \gamma_{A_i} \eta_{T_{A_i}}}, \tag{S86a}$$

$$V_B^{(M)} = \frac{\sum_{i=1}^{m} p_{S_B}^{(ith)} \gamma_{B_i} \eta_{T_{B_i}} V_B^{(ith)}}{\sum_{i=1}^{m} p_{S_B}^{(ith)} \gamma_{B_i} \eta_{T_{B_i}}}. \tag{S86b}$$

Defining $\bar{f}_{S_A} = \frac{\sum_{i=1}^{m} \gamma_{A_i} \eta_{T_{A_i}}}{m}$ and $\bar{f}_{S_B} = \frac{\sum_{i=1}^{m} \gamma_{B_i} \eta_{T_{B_i}}}{m}$, so we can express the products $\gamma_{A_i} \eta_{T_{A_i}}$ and $\gamma_{B_i} \eta_{T_{B_i}}$ as

$$\gamma_{A_i} \eta_{T_{A_i}} = \bar{f}_{S_A} + \delta(\gamma_{A_i} \eta_{T_{A_i}}), \tag{S87a}$$

$$\gamma_{B_i} \eta_{T_{B_i}} = \bar{f}_{S_B} + \delta(\gamma_{B_i} \eta_{T_{B_i}}), \tag{S87b}$$

respectively, where, $\sum_{i=1}^{m} \delta(\gamma_{A_i} \eta_{T_{A_i}}) = 0$ and $\sum_{i=1}^{m} \delta(\gamma_{B_i} \eta_{T_{B_i}}) = 0$. Substituting the Eq. (S87) into the Eq. (S86), we obtain

$$V_A^{(M)} = \frac{\sum_{i=1}^{m} p_{S_A}^{(ith)} V_A^{(ith)} \bar{f}_{S_A} + \sum_{i=1}^{m} p_{S_A}^{(ith)} V_A^{(ith)} \delta\left(\gamma_{A_i} \eta_{T_{A_i}}\right)}{\sum_{i=1}^{m} p_{S_A}^{(ith)} \bar{f}_{S_A} + \sum_{i=1}^{m} p_{S_A}^{(ith)} \delta(\gamma_{A_i} \eta_{T_{A_i}})}, \tag{S88a}$$

$$V_B^{(M)} = \frac{\sum_{i=1}^{m} p_{S_B}^{(ith)} \overline{f}_{S_B} V_B^{(ith)} + \sum_{i=1}^{m} p_{S_B}^{(ith)} V_B^{(ith)} \delta\left(\gamma_{B_i} \eta_{T_{B_i}}\right)}{\sum_{i=1}^{m} p_{S_B}^{(ith)} \overline{f}_{S_B} + \sum_{i=1}^{m} p_{S_B}^{(ith)} \delta(\gamma_{B_i} \eta_{T_{B_i}})}. \tag{S88b}$$

Assuming that the relative deviations $\delta(\gamma_{A_i} \eta_{T_{A_i}})/\overline{f}_{S_A}$ is far less than 1, we may rewrite the above equations as

$$V_A^{(M)} \approx \frac{\sum_{i=1}^{m} p_{S_A}^{(ith)} V_A^{(ith)}}{\sum_{i=1}^{m} p_{S_A}^{(ith)}} = \sum_{i=1}^{m} P_{S_A}^{(ith)} V_A^{(ith)}, \tag{S89a}$$

$$V_B^{(M)} \approx \frac{\sum_{i=1}^{m} p_{S_B}^{(ith)} V_B^{(ith)}}{\sum_{i=1}^{m} p_{S_B}^{(ith)}} = \sum_{i=1}^{m} P_{S_B}^{(ith)} V_B^{(ith)}, \tag{S89b}$$

where $P_{S_A}^{(ith)} = \frac{p_{S_A}^{(ith)}}{\sum_{i=1}^{m} p_{S_A}^{(ith)}}$ ($P_{S_B}^{(ith)} = \frac{p_{S_B}^{(ith)}}{\sum_{i=1}^{m} p_{S_B}^{(ith)}}$) is the normalized detection probability of the Stokes photon from the $i$th source in the ensemble A (B). We then express the $P_{S_A}^{(ith)} V_A^{(ith)}$ and $P_{S_B}^{(ith)} V_B^{(ith)}$ as

$$P_{S_A}^{(ith)} V_A^{(ith)} = V_A^{(M)}/m + \delta_A^{(ith)}, \tag{S90a}$$

$$P_{S_B}^{(ith)} V_B^{(ith)} = V_B^{(M)}/m + \delta_B^{(ith)}. \tag{S90b}$$

According to the Eq. (S89), we have $\sum_{i=1}^{m} \delta_A^{(ith)} = \sum_{i=1}^{m} \delta_B^{(ith)} = 0$. The Eq. (S82) can be rewritten as

$$V_{AB}^{(M)} = N \sum_{i=1}^{m} P_{S_A}^{(ith)} V_A^{(ith)} P_{S_B}^{(ith)} V_B^{(ith)}, \tag{S91}$$

where $N = \frac{\sum_{i=1}^{m} p_{S_A}^{(ith)} \sum_{i=1}^{m} p_{S_B}^{(ith)}}{\sum_{i=1}^{m} p_{S_A}^{(ith)} p_{S_B}^{(ith)}} \approx m$. Substituting the Eq. (S90) into the Eq. (S91), we obtain

$$V_{AB}^{(M)} = V_A^{(M)} V_B^{(M)} + m \sum_{i=1}^{m} \delta_A^{(ith)} \delta_B^{(ith)}, \tag{S92}$$

where $\frac{N}{m} = \frac{\sum_{i=1}^{m} p_{S_A}^{(ith)} \sum_{i=1}^{m} p_{S_B}^{(ith)}}{m \sum_{i=1}^{m} p_{S_A}^{(ith)} p_{S_B}^{(ith)}} \approx 1$ is the normalized constant. Assuming that the relative deviation $\delta_A^{(ith)} / \left( P_{S_B}^{(ith)} V_B^{(ith)} \right)$ is far less than 1, we may neglect the term $m \sum_{i=1}^{m} \delta_A^{(ith)} \delta_B^{(ith)}$ and then we can express the visibility of the state $\rho_{ent}^{(M)}$ as

$$V_{AB}^{(M)} \approx V_A^{(M)} V_B^{(M)}, \tag{S93}$$

which shows the relation between the visibility $V_{AB}^{(M)}$ of the state $\rho_{ent}^{(M)}$ and the compositive visibilities $V_A^{(M)}$ ($V_B^{(M)}$) of the multiplexed interface A (B).

Next, we define the fidelity of the entangled state $\rho_{ent}^{(M)}$ between the ensembles A and B as,

$$F_{AB}^{(M)} = \sum_{i=1}^{m} P_{S_{L0}}^{(M, ith)} F_{AB}^{(M, ith)}, \tag{S94}$$

where, $F_{AB}^{(M, ith)}$ represents the fidelity of the $i$th entangled state $\rho_{ent}^{(M, ith)}$, which is equal to that of the $i$th entangled state $\rho_{ent}^{(ith)}$, i.e., $F_{AB}^{(M, ith)} = F_{AB}^{(ith)}$. Next, we define the compositive fidelity $F_A^{(M)}$ ($F_B^{(M)}$) of the atom-photon entangled states created in the multiplexed interface A (B) as

$$F_A^{(M)} = \sum_{i=1}^{m} P_{S_A, T_A}^{(M, ith)} F_A^{(M, ith)}, \tag{S95a}$$

$$F_B^{(M)} = \sum_{i=1}^{m} P_{S_B, T_B}^{(M, ith)} F_B^{(M, ith)}, \tag{S95b}$$

respectively, where $F_{A(B)}^{(M, ith)} = \left( Tr \sqrt{ \sqrt{\rho_{r, A(B)}^{(M, ith)}} \rho_d \sqrt{\rho_{r, A(B)}^{(M, ith)}} } \right)^2$, $\rho_{r, A(B)}^{(M, ith)}$ is the reconstructed density matrix of the entangled state between the Stokes photon $S_{A_i}$ ($S_{B_i}$) and anti-Stokes photon $T_{A_i}$ ($T_{B_i}$) from the source $i$ in the multiplexed interface A (B), $\rho_d$ is the density matrix of the ideal entangled state $|\Phi\rangle_{S,T}^{'(ith)} = \left( \cos\vartheta |H\rangle_{T_i} |H\rangle_{S_i} + \sin\vartheta |V\rangle_{T_i} |V\rangle_{S_i} \right)$.

We then discuss the dependence of the fidelity $F_{AB}^{(M)}$ on the compositive fidelities $F_A^{(M)}$ and $F_B^{(M)}$. For an entangled state, the relation between the fidelity $F$ and the visibility $V$ of the state is given by [12] $F = (3V+1)/4$. So, we can write

$$F_{AB}^{(M, ith)} = \frac{3V_{AB}^{(M, ith)} + 1}{4}, \tag{96a}$$

$$F_A^{(M, ith)} = \frac{3V_A^{(M, ith)} + 1}{4}, \tag{96b}$$

$$F_B^{(M, ith)} = \frac{3V_B^{(M, ith)} + 1}{4}. \tag{96c}$$

Based on the Eqs. (78, 83, 94-96), we then obtain

$$F_{AB}^{(M)} = \sum_{i=1}^{m} P_{S_{L0}}^{(M,ith)} F_{AB}^{(M, ith)} = \sum_{i=1}^{m} \left[ P_{S_{L0}}^{(M,ith)} \cdot \frac{3V_{AB}^{(M, ith)} + 1}{4} \right] = \frac{3V_{AB}^{(M)} + 1}{4}, \tag{S97a}$$

$$F_A^{(M)} = \sum_{i=1}^{m} P_{S_A,T_A}^{(M,ith)} F_A^{(M, ith)} = \sum_{i=1}^{m} \left[ P_{S_A,T_A}^{(M,ith)} \cdot \frac{3V_A^{(M, ith)} + 1}{4} \right] = \frac{3V_A^{(M)} + 1}{4}, \tag{S97b}$$

$$F_B^{(M)} = \sum_{i=1}^{m} P_{S_B,T_B}^{(M,ith)} F_B^{(M, ith)} = \sum_{i=1}^{m} \left[ P_{S_B,T_B}^{(M,ith)} \cdot \frac{3V_B^{(M, ith)} + 1}{4} \right] = \frac{3V_B^{(M)} + 1}{4}. \tag{S97b}$$

According to the above equations and the relation $V_{AB}^{(M)} \approx V_A^{(M)} V_B^{(M)}$ of Eq. (S93), we obtain

$$F_{AB}^{(M)} \approx \left(1 + (4F_A^{(M)} - 1)(4F_B^{(M)} - 1)/3\right)/4, \tag{S98}$$

which shows the relation between the fidelity $F_{AB}^{(M)}$ of the state $\rho_{ent}^{(M)}$ and the compositive fidelities $F_A^{(M)}$ ($F_B^{(M)}$) of the multiplexed interface A ( B).

Subsequently, we define the Bell parameter for the entangled state $\rho_{ent}^{(M)}$, which is given by

$$S_{AB}^{(M)} = \sum_{i=1}^{m} P_{S_{L0}}^{(M, ith)} S_{AB}^{(M, ith)}, \tag{S99}$$

where, $S_{AB}^{(M, ith)}$ is the Bell parameter for the two photons $T_{A_i}$ and $T_{B_i}$ retrieved from ensembles A and B, respectively. Then, we define the compositive Bell parameters for the multiplexed interfaces A and B, respectively, which are written as

$$S_A^{(M)} = \sum_{i=1}^{m} P_{S_A,T_A}^{(M, ith)} S_A^{(M, ith)},  \quad (S100a)$$

$$S_B^{(M)} = \sum_{i=1}^{m} P_{S_B,T_B}^{(M, ith)} S_B^{(M, ith)},  \quad (S100b)$$

where, $S_A^{(M, ith)}$ ($S_B^{(M, ith)}$) is the Bell parameter for the Stokes and anti-Stokes photons from the source $i$ in the ensemble A (B) for the multiplexed case. Utilizing the relations [13] $S_{AB}^{(M, ith)} = 2\sqrt{2} V_{AB}^{(M, ith)}$, $S_{A(B)}^{(M, ith)} = 2\sqrt{2} V_{A(B)}^{(M, ith)}$, Eq. (S99) and Eq. (S100), we obtain

$$S_{AB}^{(M)} = \sum_{i=1}^{m} P_{S_{L0}}^{(M, ith)} S_{AB}^{(M, ith)} = \sqrt{2} \sum_{i=1}^{m} P_{S_{L0}}^{(M,ith)} \cdot V_{AB}^{(M, ith)} = \sqrt{2} V_{AB}^{(M)},  \quad (S101a)$$

$$S_A^{(M)} = \sum_{i=1}^{m} P_{S_A,T_A}^{(M,ith)} S_A^{(M, ith)} = \sqrt{2} \sum_{i=1}^{m} P_{S_A,T_A}^{(M,ith)} \cdot V_A^{(M, ith)} = \sqrt{2} V_A^{(M)},  \quad (S101b)$$

$$S_B^{(M)} = \sum_{i=1}^{m} P_{S_B,T_B}^{(M,ith)} S_B^{(M, ith)} = \sum_{i=1}^{m} P_{S_B,T_B}^{(M,ith)} V_B^{(M, ith)} = \sqrt{2} V_B^{(M)}.  \quad (S101c)$$

Based on the relation $V_{AB}^{(M)} \approx V_A^{(M)} V_B^{(M)}$ of Eq. (S93), we obtain

$$S_{AB}^{(M)} = S_A^{(M)} S_B^{(M)} / 2\sqrt{2},  \quad (S102)$$

which shows the relation between the Bell parameter $F_{AB}^{(M)}$ of the state $\rho_{ent}^{(M)}$ and the compositive Bell parameters $S_A^{(M)}$ ($S_B^{(M)}$) of the multiplexed interface A (B).

For one of the two ensembles (A and B ensembles), we can write the Bell parameter $S^{(ith)}$ for the two entangled photon from the $i$th SWPE source $|\Phi\rangle^{(ith)}$ in this ensemble as

$$S^{(ith)} = \left| E^{(ith)}(\theta_{S_i}, \theta_T) - E^{(ith)}(\theta_{S_i}, \theta_T') + E^{(ith)}(\theta_{S_i}', \theta_T) + E^{(ith)}(\theta_{S_i}', \theta_T') \right| < 2,  \quad (S103)$$

where $\theta_{S_i}$ and $\theta_T$ is the polarization angle of the Stokes photon $S_i$ and the anti-Stokes photon $T_i$, which are set by using two $\lambda/2$ plates, respectively. The correlation function $E^{(ith)}(\theta_{S_i}, \theta_T)$ is defined by

$$E^{(ith)}(\theta_{S_i}, \theta_T) = \frac{C_{HH}^{(ith)}(\theta_{S_i}, \theta_T) + C_{VV}^{(ith)}(\theta_{S_i}, \theta_T) - C_{HV}^{(ith)}(\theta_{S_i}, \theta_T) - C_{VH}^{(ith)}(\theta_{S_i}, \theta_T)}{C_{HH}^{(ith)}(\theta_{S_i}, \theta_T) + C_{VV}^{(ith)}(\theta_{S_i}, \theta_T) + C_{HV}^{(ith)}(\theta_{S_i}, \theta_T) + C_{VH}^{(ith)}(\theta_{S_i}, \theta_T)}, \quad (S104)$$

where, for example, $C_{HH}(\theta_{S_i}, \theta_T)$ is the coincidence detection rate between the detectors $D_{S_i}$ and $D_T$ for the polarization angle $\theta_{S_i}$ and $\theta_T$.

The compositive Bell parameter for the multiplexed interface with $m$ SWPE sources generated from the ensemble can be rewritten as

$$S^{(M)} = \sum_{i=1}^{m} P_{S,T}^{(M, ith)} S^{(M, ith)} = \left| E^{(M)}(\theta_{S_i}, \theta_T) - E^{(M)}(\theta_{S_i}, \theta_T') + E^{(M)}(\theta_{S_i}', \theta_T) + E^{(M)}(\theta_{S_i}', \theta_T') \right| < 2, \quad (S105)$$

where, $S^{(M, ith)}$ is the Bell parameter for the multiplexed case, with $S^{(M, ith)} = S^{(ith)}$, $E^{(M)}(\theta_{S_i}, \theta_T) = \sum_{i=1}^{m} P_{S,T}^{(M, ith)} E^{(M, ith)}(\theta_{S_i}, \theta_T)$ is the correlation function for the multiplexed interface, $P_{S,T}^{(M, ith)}$ is the normalized coincidence probability, which is defined as

$$P_{S,T}^{(M, ith)} = \frac{C_{HH}^{(M, ith)}(\theta_{S_i}, \theta_T) + C_{VV}^{(M, ith)}(\theta_{S_i}, \theta_T) + C_{HV}^{(M, ith)}(\theta_{S_i}, \theta_T) + C_{VH}^{(M, ith)}(\theta_{S_i}, \theta_T)}{\sum_{i=1}^{m} \left( C_{HH}^{(M, ith)}(\theta_{S_i}, \theta_T) + C_{VV}^{(M, ith)}(\theta_{S_i}, \theta_T) + C_{HV}^{(M, ith)}(\theta_{S_i}, \theta_T) + C_{VH}^{(M, ith)}(\theta_{S_i}, \theta_T) \right)}. \quad (S106)$$

For example, $C_{SH,TH}^{(M, ith)}(\theta_{S_i}, \theta_T)$ ($C_{SV,TV}^{(M, ith)}(\theta_{S_i}, \theta_T)$) is the coincidence detection rate between the detectors $D_{S_i}^{(1)}$ ($D_{S_i}^{(2)}$) and $D_T^{(1)}$ ($D_T^{(2)}$) for the polarization angle $\theta_{S_i}$ and $\theta_T$ for the multiplexed case. According to the above Eqs. (S105) and (S106), we obtain:

$$E^{(M)}(\theta_{S_i}, \theta_T) = \frac{\sum_{i=1}^{m} \left( C_{SH,TH}^{(M, ith)}(\theta_{S_i}, \theta_T) + C_{SV,TV}^{(M, ith)}(\theta_{S_i}, \theta_T) - C_{SH,TV}^{(M, ith)}(\theta_{S_i}, \theta_T) - C_{SV,TH}^{(M, ith)}(\theta_{S_i}, \theta_T) \right)}{\sum_{i=1}^{m} \left( C_{SH,TH}^{(M, ith)}(\theta_{S_i}, \theta_T) + C_{SV,TV}^{(M, ith)}(\theta_{S_i}, \theta_T) + C_{SH,TV}^{(M, ith)}(\theta_{S_i}, \theta_T) + C_{SV,TH}^{(M, ith)}(\theta_{S_i}, \theta_T) \right)}. \quad (S107)$$

We now discuss the difference between the compositive Bell parameter $S^{(M)}$ and the average Bell parameter for the $m$-SWPE sources defined by

$$\bar{S} = \sum_{i=1}^{m} S^{(M, ith)} / m. \quad (S108)$$

Based on the definition, we have $S^{(M, ith)} = \overline{S} + \delta S^{(M, ith)}$ and $\sum_{i=1}^{m} \delta S^{(M, ith)} = 0$. We then define the average coincidence probability

$$\overline{P_{S,T}} = \frac{\sum_{i=1}^{m} P_{S,T}^{(M, ith)}}{m} = \frac{1}{m}. \tag{S109}$$

In this case, we have $P_{S,T}^{(M, ith)} = \overline{P_{S,T}} + \delta P_{S,T}^{(M, ith)}$ and $\sum_{i=1}^{m} \delta P_{S,T}^{(M, ith)} = 0$. So, we obtain

$$S^{(M)} = \sum_{i=1}^{m} P_{S,T}^{(M, ith)} S^{(M, ith)} = \sum_{i=1}^{m} \left( \overline{P_{S,T}} + \delta P_{S,T}^{(M, ith)} \right) \left( \overline{S} + \delta S^{(M, ith)} \right). \tag{S110}$$

When both deviations $\delta P_{S,T}^{(M, ith)}$ and $\delta S^{(M, ith)}$ are far less than $\overline{P_{S,T}}$ and $\overline{S}$, respectively, their products $\delta P_{S,T}^{(M, ith)} \delta S^{(M, ith)}$ can be neglected in the Eq. (S105) and then we obtain

$$S^{(M)} \approx \overline{S}. \tag{S111}$$

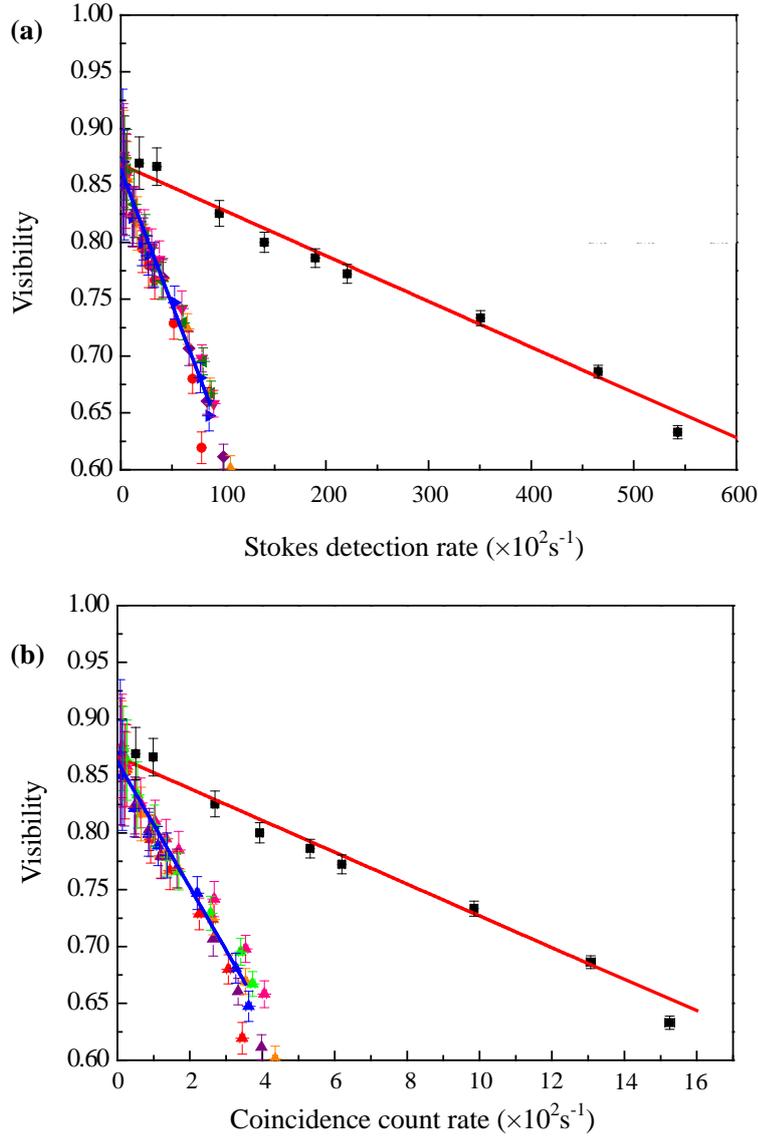

Supplementary Figure S1 (color online). The polarization visibility as the functions of the Stokes detection rate (a) and the coincidence count rate (b) for the polarization setting of *D-A*, respectively. The red, yellow, pink, green, blue and purple data are the measured polarization visibilities $V_{D-A}^{(1st)}$, $V_{D-A}^{(2nd)}$, …and $V_{D-A}^{(6th)}$ as the functions of the Stokes detection (coincidence count) rates $R_{D-A}^{(1st)}$, $R_{D-A}^{(2nd)}$, … $R_{D-A}^{(6th)}$ ($C_{D-A}^{(1st)}$, $C_{D-A}^{(2nd)}$, … $C_{D-A}^{(6th)}$) of the individual sources 1, 2, …, 6, respectively, which are measured under the non-multiplexed case (without optical switching network). The black data in the (a) and (b) are the measured visibilities $V_{D-A}^{(M)}$ for the multiplexed interface versus the total Stokes detection $R_{D-A}^{(m)}$ and the total coincidence count rate $C_{D-A}^{(m)}$, respectively. The blue solid lines in the (a) and (b) are the least-square fittings to the single-source data according to the Eqs. (S48) and (S49), respectively, the red solid lines in the (a) and (b) are the least-square fittings to the $V_{D-A}^{(M)}$ according to the Eqs. (S57) and (S59), respectively.

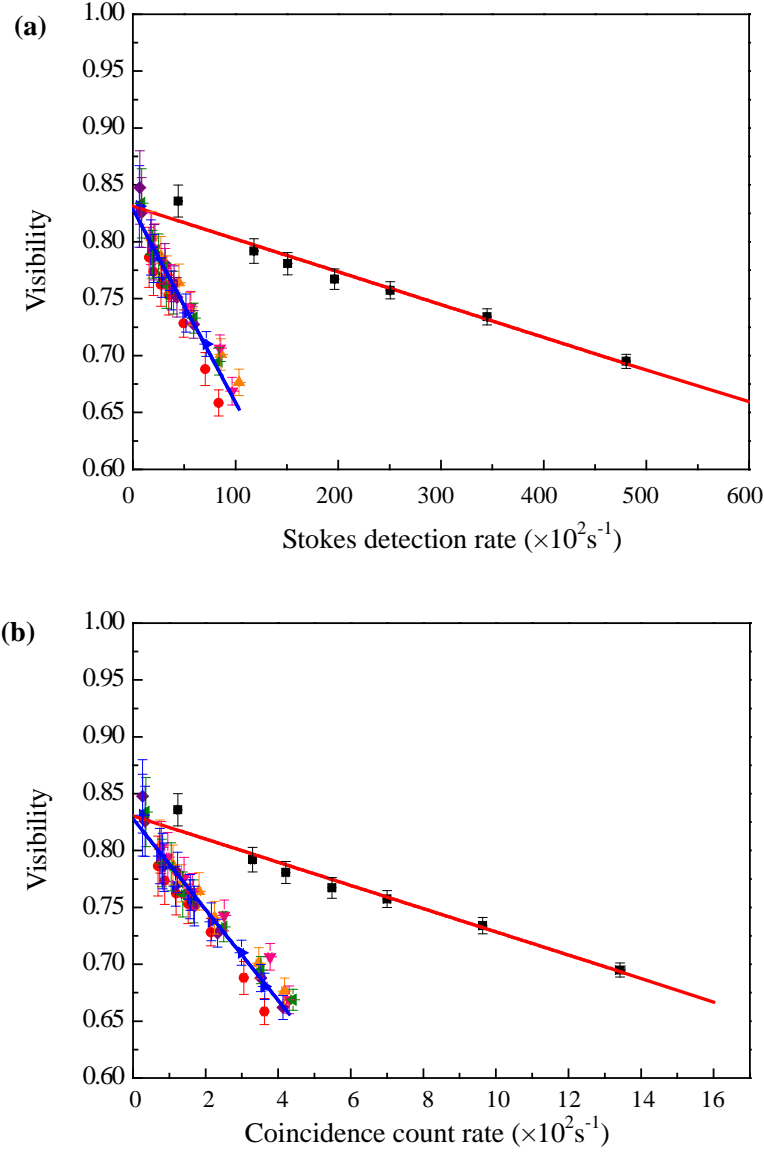

Supplementary Figure S2 (color online). The polarization visibility as functions of Stokes detection rate (a) and coincidence count rate (b) for the polarization setting of *R-L*, respectively. The red, yellow, pink, green, blue and purple data are the measured polarization visibilities $V_{R-L}^{(1st)}$, $V_{R-L}^{(2nd)}$, …and $V_{R-L}^{(6th)}$ as the functions of Stokes detection (coincidence count) rates $R_{R-L}^{(1st)}$, $R_{R-L}^{(2nd)}$, … $R_{R-L}^{(6th)}$ ( $C_{R-L}^{(1st)}$, $C_{R-L}^{(2nd)}$, … $C_{R-L}^{(6th)}$ ) for the individual sources 1, 2, …, 6, respectively, which are measured under the non-multiplexed case (without optical switching network). The black data in the (a) and (b) are the measured visibilities $V_{R-L}^{(M)}$ for the multiplexed interface versus total Stokes detection $R_{R-L}^{(m)}$ (a) and total coincidence count rate $C_{R-L}^{(m)}$ (b), respectively. The blue solid lines in the (a) and (b) are the least-square fittings to the single-source data according to the Eqs. (S48), and (S49), respectively, the red solid lines in the (a) and (b) are the least-square fittings to the $V_{R-L}^{(M)}$ according to the Eqs. (S57) and (S59), respectively.

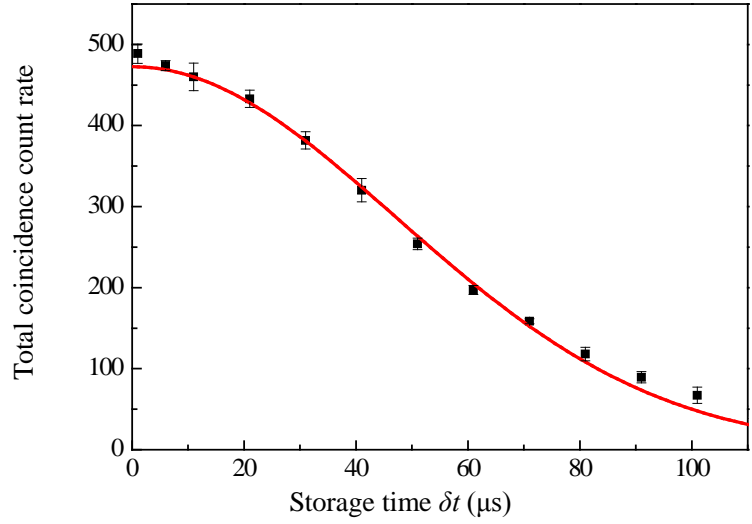

Supplementary Figure S3 (color online). Total coincidence count rate $C_{H-V}^{(6)}$ as a function of storage time $\delta t$ for the write-pulse power fixed to get $p_S^{(6)} \approx 0.0297$. The red solid curve is a fitting basing on the form $C_{H-V}^{(m)}(\delta t) = C_0 \exp(-\delta t^2 / \tau^2)$ yielding a lifetime of $\tau = 66.7\,\mu s$.

|  | $i=1$ | $i=2$ | $i=3$ | $i=4$ | $i=5$ | $i=6$ |
|---|---|---|---|---|---|---|
| $\eta_{OSN_i}$ | 0.83 | 0.85 | 0.86 | 0.84 | 0.85 | 0.84 |
| $\eta_{CF_i}$ | 0.83 | 0.79 | 0.82 | 0.82 | 0.8 | 0.79 |
| $\eta_{RC_i}$ | 0.689 | 0.672 | 0.705 | 0.689 | 0.680 | 0.664 |

Supplementary Table S1. The measured optical switching network transmission $\eta_{OSN_i}$, CSMF coupling efficiency $\eta_{CF_i}$ and transmission of the router circuitry $\eta_{RC_i}$, with $\eta_{RC_i} = \eta_{OSN_i} \cdot \eta_{CF_i}$.

|  | $i=1$ | $i=2$ | $i=3$ | $i=4$ | $i=5$ | $i=6$ |
|---|---|---|---|---|---|---|
| $\eta_{Filter_{S_i}}$ | 0.78 | 0.79 | 0.78 | 0.80 | 0.79 | 0.77 |
| $\eta_{SMF_{Si}}$ | 0.79 | 0.78 | 0.80 | 0.78 | 0.80 | 0.79 |
| $\eta_{S_i}$ | 0.29 | 0.29 | 0.29 | 0.30 | 0.30 | 0.29 |

Supplementary Table S2. The measured efficiencies of the optical filters $\eta_{Filter_{S_i}}$, the coupling efficiency of fiber coupler $\eta_{SMF_{Si}}$ and total detection efficiencies $\eta_{S_i}$, for the Stokes optical field mode $S_i$.

|  | $i=1$ | $i=2$ | $i=3$ | $i=4$ | $i=5$ | $i=6$ |
|---|---|---|---|---|---|---|
| $\eta_{Filter_{T_i}}$ | 0.81 | 0.79 | 0.80 | 0.79 | 0.79 | 0.78 |
| $\eta_{SMF_{T_i}}$ | 0.78 | 0.77 | 0.78 | 0.80 | 0.78 | 0.76 |
| $\eta_{T_i}$ | 0.30 | 0.29 | 0.29 | 0.30 | 0.28 | 0.29 |

Supplementary Table S3. The measured efficiencies of the optical filters $\eta_{Filter_{T_i}}$, the coupling efficiency of fiber coupler $\eta_{SMF_{T_i}}$, detection efficiencies $\eta_{T_i}$ for the anti-Stokes optical field mode $T_i$.